\documentclass[structabstract]{aa} % for a referee version
\usepackage{graphicx}
\newcommand{\lia}{$^7$Li}
\newcommand{\lib}{$^6$Li}
\newcommand{\lis}{$^6$Li/$^7$Li }
\newcommand{\lir}{$^7$Li/$^6$Li }
\newcommand{\bb}{$^{11}$B/$^{10}$B }

\newcommand{\ms}{M$_{\odot}$}
\newcommand{\zs}{Z$_{\odot}$}

\newcommand{\mex}{$M_{Exp}$}

\newcommand{\mej}{$M_{Ej}$}

\newcommand{\neb}{$^{22}$Ne}

\newcommand{\neo}{$^{22}$Ne/$^{20}$Ne}

\begin{document}
   \title{Production and evolution of Li, Be and B isotopes in the Galaxy}

   \author{N.Prantzos \inst{1}   }

   \institute{Institut d'Astrophysique de Paris, UMR7095 CNRS, Univ.P. \& M.Curie, 98bis Bd. Arago, 75104 Paris, France;
         \email{prantzos@iap.fr}
             }
   \date{}

% \abstract{{{{{} 
% 5 {} token are mandatory
 
  \abstract
   {We reassess the problem of the production and evolution of the light elements
Li, Be and B and of their isotopes in the Milky Way in the light of new observational and theoretical
developments. }
{The main novelty is the introduction of a new scheme for the origin
of Galactic cosmic rays (GCR), which for the first time enables a self-consistent calculation of their 
composition during galactic evolution.  }
{The scheme accounts for key features of the present-day GCR source   composition, it is based
on the wind yields of the Geneva models of rotating, mass-losing stars and it is fully coupled to a detailed
 galactic chemical evolution code. }
   {We find that the adopted GCR source composition accounts naturally for the observations of primary Be
   and helps understanding why Be follows Fe more closely than O. We find that GCR
   produce $\sim$70\% of the solar \bb \ isotopic ratio; the remaining 30\% of $^{11}$B 
   presumably result from $\nu$-nucleosynthesis in massive star explosions. We find that GCR and
   primordial nucleosynthesis can produce at  most $\sim$30\% of solar Li. At least half of the  
   solar Li has to originate in low-mass stellar sources (red giants, asymptotic giant branch stars, or novae),
   but the required average yields of those sources are found to be much higher than obtained in current 
   models of stellar nucleosynthesis. 
   We also present radial profiles of LiBeB elemental and isotopic 
   abundances in the Milky Way disc. We argue that
   the shape of those profiles - and the late evolution of LiBeB in general -  
   reveals important features of the production of those light elements through
   primary and secondary processes.  }
   {}


   \maketitle

\section{Introduction}
The idea that the light and fragile elements Li, Be and B are
produced by the interaction  of the energetic nuclei
of galactic cosmic rays (CGR) with the nuclei of the interstellar medium
(ISM) was introduced 40 years ago (Reeves et al. 1970; Meneguzzi et al. 1971, hereafter MAR).
In those early works it was shown that taking into account
the relevant cross-sections and with plausible assumptions about the
GCR properties - source composition, intensity, and spectrum -
one may reproduce  the abundances of those light
elements observed in GCR and in  meteorites (=pre-solar) reasonably well. The only exception is Li, which can
have only a minor contribution ($<$20\%) from GCR and requires a stellar source. Despite
more than 30 years of theoretical and observational work, the stellar source of Li
remains elusive at present.

A new impetus was given to the subject  by observations of halo stars in the 1990ies showing
that Be and B behave as Fe, i.e. as  primary elements (Gilmore
et al. 1992; Ryan  et al. 1992; Duncan et al. 1992),  contrary to theoretical expectations.
The reason for this "puzzling" behaviour was rapidly inferred by Duncan et al. (1992):
GCR must have a metallicity-independent composition  to produce primary LiBeB (see also Prantzos 1993).
Other ideas (e.g. Prantzos et al. 1993) were only partially successful in that respect (see Reeves 1994
for a summary of the situation in the mid-90ies).
Ramaty et al. (1997) showed that a metallicity-independent GCR 
composition is the only viable alternative for energetic reasons: if in the early Galaxy
GCR had a metallic content much lower than today, they would need much more energy than 
supernovae can provide  to always yield primary Be. 
It was claimed that GCR can aquire a metallicity-independent composition in the environment of
superbubbles, powered and enriched by the ejecta of dozens of massive stars and supernovae (Higdon et al. 1997).
In the absence of convincing alternatives, the "superbubble paradigm" became  the physical
explanation for both the origin of GCR (e.g. Parizot et al. 2004)
and - by default - for primary Be (despite some criticism,
e.g. in Prantzos 2006a).

Independently of the crucial question of the GCR origin, the Be and B observations of the 1990ies
made it necessary to link the physics of GCR to detailed models of galactic chemical evolution
(Prantzos et al. 1993;  Ramaty et al. 1997).
In the past few years, important developments occured in both observations and theory,
making  a reassessment of this vast subject necessary. 

From the observational side, large surveys of Be in stars of low metallicities 
(Primas 2010; Tan et al. 2009; Smiljianic et al. 2009;  
Boesgaard et al. 2011) considerably improved the statistics of the Be vs Fe, 
but also of Be vs. O relationships,
providing combined and  tighter constraints to models than those previously available. 
Furthermore, observations of  Li isotopic ratios became available, both in low-metallicity
halo stars (Asplund et al. 2006; Garcia-Perez et al. 2009)  and in the local ISM 
(Kawanomoto et al. 2009). The former, suggesting a surprisingly high  \lis \ ratio
in the early Galaxy, stimulated a large body of theoretical work (see Prantzos 2006b 
and references therein) but remains  controversial (Spite and Spite 2010 and references therein); 
the latter, combined to the well-known meteoritic ratio of \lis, constrains 
the late evolution of Li isotopes in the local region of the Galaxy.

On the theoretical side, Prantzos (2012) 
argued that GCR are accelerated mainly by the forward shocks 
of supernova explosions, propagating through the winds of massive stars and  the ISM.
By using detailed recent models of nucleosynthesis in massive, mass losing stars, 
he showed   quantitatively that the most prominent feature of the observed  GCR composition,
namely the high isotopic \neo \ ratio ($\sim$5 times solar),  can be nicely
obtained if acceleration occurs in the early Sedov phase of supernova remnants, for
shock velocities $>$1500 km/s. Furthermore, models of rotating, mass-losing stars 
were calculated  by the Geneva group (Hirschi et al. 2005), 
showing that the amounts of CNO nuclei  released in the stellar winds  are almost independent of
stellar metallicity (Hirschi 2006). These  theoretical results open the  way, for the first time, 
for a self-consistent calculation 
of the GCR composition (and the resulting LiBeB production from spallation of CNO) 
throughout the whole Galactic evolution.  

The aim of this work is threefold: 1) to evaluate the GCR (spallogenic) production of LiBeB
on the basis of the new scheme for the GCR origin;
2) to constrain the yields of the stellar sources of Li by removing the contribution
of GCR and primordial nucleosynthesis; 3) to explore the late evolution of the Li and B isotopic ratios,
both in the solar neighbourhood and in the Galactic disc, 
 to assess the importance of the secondary component of their production 
 (which reveals itself only at high metallicities).
The plan of the paper is as follows:

In Sec. 2 we present an overview of the problem of LiBeB production   by GCR. After a short
 presentation of some basic results (Sec. 2.1), we discuss the
problem raised by the observed primary behaviour of Be vs. Fe (Sec. 2.2) and of the implications 
it has for the composition of GCR. By comparing the various ideas for the origin of GCR
we conclude that the only site compatible with all direct observational requirements
(which concern the  {\it present-day} GCR composition)
is the one involving shock waves propagating through the winds of massive stars (Sec. 2.3).
We argue then, based on recent stellar models (Sec. 2.4), 
that the circumstellar environment of {\it rotating}, mass-losing stars 
naturally provides a GCR composition across Galactic history that is compatible with the 
observed evolution of Be. 

In Sec. 3.1 we present  the model for the chemical evolution of the Milky Way (stellar initial mass function, supernova rates, yields of chemical elements, observational constraints other than those of LiBeB), 
and the  results obtained with this model 
for the evolution of the key elements C, N and O in the ISM. In Sec. 3.2 we calculate the
composition of the GCR as a  function of metallicity, by using the results of the chemical evolution model
and the new scheme of the GCR origin presented in Sec. 2.  In Sec. 3.3 we present in some detail the
adopted treatment of the LiBeB production from GCR (spectra, composition, coupling to SN energetics, etc.).

In Sec. 4.1 we  discuss the results obtained for Be and the reason for its primary behaviour with respect to Fe
rather than O. In Sec. 4.2  we present results for the B isotopes, 
in particular  concerning the contribution  of neutrino nucleosynthesis in core-collapse supernovae (CCSN) to $^{11}$B production.
 
The evolution of the Li isotopes is considered in Sec. 5. In Sec. 5.1 we consider 
the evolution of \lia \ and from our models we infer the required 
yields of the low-mass stellar component of that isotope,
after taking into account contributions from primordial nucleosynthesis and GCR. 
We find that the required \lia \
yields are much higher than those calculated in the literature. We discuss the
controversial question of high early \lib \ in Sec. 5.2 and the implications of the late evolution of 
\lis \ ratio in Sec. 5.3. In Sec. 6 we discuss 
the radial profiles of LiBeB and of the corresponding isotopic
ratios across the Milky Way disc. Summary and conclusions are presented in Sec. 7.

\section{Overview of the subject}
\label{Sec:Overview}

\subsection{Definitions and basic results}
\label{Subsec:Formalism}

The present-day abundances of LiBeB, produced after $\sim$10 Gyr of cosmic evolution through spallation of CNO nuclei 
by GCR, can be obtained in a straightforward way, at least to a first approximation\footnote{The
full calculation should include production by spallation  of other primary and secondary 
nuclides, such as $^{13}$C; however, this has only second order effects.}. 
The production rate (s$^{-1}$)
of the abundance $Y_L=N_L/N_H$ (per H atom) of LiBeB nuclei in the ISM is given by
\begin{eqnarray}
\frac{dY^{ISM}_L}{dt} \ &=& \ F^{GCR}_{p,a}\sigma_{pa+CNO}Y^{ISM}_{CNO}   \nonumber \\ 
                  &+& \ F^{GCR}_{CNO}\sigma_{pa+CNO}Y^{ISM}_{p,a} P_L \nonumber \\  
                  &+& \ F^{GCR}_{a}\sigma_{a+a}Y^{ISM}_{a} P_L,
                  \label{eqsimple}
\end{eqnarray}
where $F$ (cm$^{-2}$ s$^{-1}$) is the average GCR flux of protons, alphas or CNO, $Y^{ISM}$ the abundances 
(per H atom) of those nuclei
in the ISM, and $\sigma$ (cm$^2$) is the average (over the equilibrium energy spectrum of GCR) 
cross-section for the corresponding spallation reactions producing LiBeB.
The first term of the right-hand side of this 
 equation (fast protons and alphas hitting CNO nuclei of the ISM) is
known as the {\it direct} term, the second one (fast CNO nuclei being fragmented on ISM protons and 
alphas) is the {\it reverse} term, and the last one involves ``spallation-fusion" $\alpha + \alpha$
reactions, concerning only the Li isotopes. 
$P_L$ is the probability that nuclide $L$ (produced at high energy) 
will be thermalised and remain in the ISM (see Sec. 3.3 for details). 
Obviously, the GCR flux term $F^{GCR}_{CNO} \propto Y^{GCR}_{CNO}$
is proportional to the abundances of CNO nuclei in GCR, a fact of paramount importance for the evolution
of Be and B, as we shall see below.

In Eq. (1) one may  substitute typical  values - see MAR - for GCR fluxes ($F^{GCR}_p\sim$10 p cm$^{-2}$ s$^{-1}$ 
 for protons and scaled values for other GCR nuclei), for the corresponding cross sections (averaged
over the GCR equilibrium spectrum $\sigma_{p,a+CNO\longrightarrow Be}\sim$10$^{-26}$ cm$^{2}$) and
for ISM abundances ($Y_{CNO}\sim$10$^{-3}$);  integrating for $\Delta t \sim$10 Gyr, one
finds then that $Y_{Be}\sim$2 10$^{-11}$, i.e. approximately the meteoritic Be value (Lodders 2003). 
Satisfactory results are also obtained for the abundances of $^6$Li  and $^{10}$B.
Despite the crude approximations adopted (constant GCR fluxes and ISM abundances for 10 Gyr,
 average production cross sections, secondary production channels ignored),
the above calculation correctly reproduces  both the absolute values (within a factor of two) and the relative
values (within 10 \% of the solar abundances) of $^6$Li, $^9$Be and $^{10}$B. This constitutes the strongest,
quantitative, argument for the validity of the idea of LiBeB produced by GCR.
 
In the original MAR paper, two problems were identified with the GCR production of LiBeB nuclei, compared
to the meteoritic composition: they concern
the \lir ratio, which is $\sim$2 in GCR, but $\sim$12 in meteorites;  and the \bb
ratio, which is $\sim$2.5 in GCR, but $\sim$4 in meteorites. It was then suggested in MAR that supplementary
sources are needed for $^7$Li and $^{11}$B. The idea of an {\it ad hoc}  low-energy GCR component (a "carrot"),
producing mostly \lia \ (because of the corresponding large $\alpha+\alpha$ cross sections at low energies) was quantitatively
explored in Menneguzzi and Reeves (1975). However, it implied high ionisation rates of the ISM near the
acceleration sites and strong $\gamma$-ray fluxes - from p-p collisions and subsequent pion decay -
which have not been detected (see, however, Indriolo and McCall 2011 for recent observations suggesting
higher ionisation rates than found before).

Modern solutions to
those problems involve {\it stellar} production of $\sim$40\% 
of $^{11}$B (through $\nu$-induced spallation of $^{12}$C in CCSN, see Sec. 4.2) and of $\sim$60\% of
$^7$Li (in the hot envelopes of red giants, AGB stars and/or novae, see Sec. 5.1). In both
cases, however, uncertainties in the yields are such that observations
are used to  constrain the yields of the candidate sources
rather than to confirm the validity of the scenario. 
We shall turn to those questions in the corresponding
sections.

\subsection{The problem of primary Be}

Observations of halo stars in the 90ies revealed a linear relationship
between Be/H and Fe/H (Gilmore et al. 1992; Ryan et al. 1992) 
as well as between B/H and Fe/H (Duncan et al. 1992). 
That was unexpected, since Be and B
were thought to be produced as {\it secondaries}\footnote{In galactic chemical evolution, {\it primary} elements
have a production rate independent of metallicity (they are produced by H and He), while {\it secondary} elements
have  a production rate proportional - explicitly or implicitly -  on metallicity (or time, because
metallicity increases approximately proportionally to time and thus  can be taken as  a proxy for it).}. 
It is clear that the "direct" term in Eq. \ref{eqsimple}  leads to 
secondary production of Be and B, since it depends explicitly on the ever increasing 
abundances $Y_{CNO}$ of the ISM.
The "reverse" term was thought to be symmetric to the direct one, with the GCR fluxes of CNO
nuclei  $F_{CNO}^{GCR}$ being proportional to the ISM abundances of those nuclei.
Indeed, according to the "paradigm" for the GCR composition in the 80ies (Meyer 1985),
GCR originate in the coronae of ordinary, low-mass stars (sharing the composition of the ambient ISM), from where
they are injected in the ISM and are  subsequently accelerated by SN shock waves.
For both the direct and the reverse terms, then, the production rate of Be in Eq. \ref{eqsimple} is
proportional to $Y_{CNO}^{ISM}$ and leads to Be production as a secondary. 
 Only the Li isotopes, produced at low 
metallicities mostly by $\alpha+\alpha$  reactions - third term in Eq. 1 -   were thought to be produced  as 
primaries (Steigman and Walker 1992), since the abundance of $^4$He varies little during
galactic evolution. However, Li abundance at low metallicities is totally
dominated by primordial $^7$Li, and the small fraction of $^6$Li was  below detectability levels
in the 90ies.

\begin{figure}
\begin{center}
\includegraphics[width=0.49\textwidth]{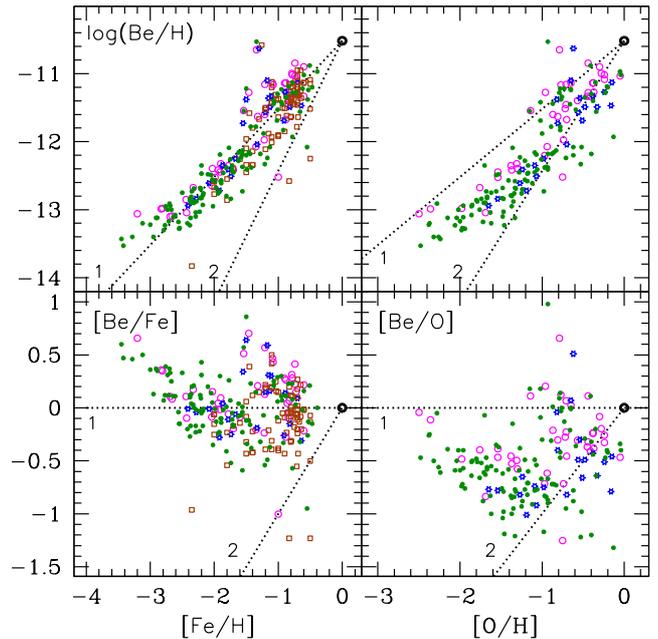}
\caption[]{ Observations of Be vs. Fe ({\it left}) and vs. O ({\it right}). In all panels, 
{\it dotted lines} indicate slopes of 1 (primary) and 2 (secondary). 
%Be clearly behaves  as a primary vs. Fe, whereas there is more scatter in the data vs. O.
Data are from Primas (2010, {\it circles}), Tan et al. (2009, {\it asterisks}), Smiljanic et al. (2009, {\it open squares}),
and Boesgaard et al. (2011, {\it dots}).
}
\label{Fig:Be-obs}
\end{center}
\end{figure}

A compilation of recent measurements for Be appears in Fig. \ref{Fig:Be-obs}, 
as a function of Fe/H and of O/H\footnote{It should be emphasized that 
the scatter displayed in Fig. \ref{Fig:Be-obs} - and all other figures displaying abundance
data from different sources - is partly due to systematic uncertainties in the data analysis (effective temperature scales,
oxygen abundance indicators, etc.); the true scatter would be less than suggested by the figure.}.
Evidently, Be/H behaves as a primary with respect to Fe in the
whole metallicity range (covering three orders of magnitude, from [Fe/H]=-3.4 to [Fe/H]=0)
while the situation with respect to O is more complex: while at low metallicities ([O/H]$<$-1)
the slope of Be/H vs O/H is 1, at higher metallicities the scatter in the data prevents one
from defining a slope; however, a secondary-like behaviour is apparently required to explain the 
late evolution of Be up to [O/H]=0. The lower panels of Fig. \ref{Fig:Be-obs}
display those same features more clearly, emphasising the large scatter of Be/Fe or Be/O 
(about 1 dex) at any metallicity.

The only way to produce primary Be  is
by assuming that GCR always have the same CNO content, as suggested in Duncan et al. (1992).
In the first ever work combining a detailed chemical evolution code with the physics of
LiBeB production by GCR, 
Prantzos et al. (1993) attempted to  enhance the early production of secondary Be by 
invoking a better confinement of GCR in the early Galaxy, leading to higher GCR fluxes $F^{GCR}$; a similar
reasoning was adopted in Fields et al. (2001).
Fields and Olive (1999) explored the possibility of a high O/Fe at low metallicities,
which  increases the contribution of the term $Y_{CNO}^{ISM}$. All those efforts - and a few others -
slightly alleviated  the problem, but could not solve it.
The reason for that failure
was clearly revealed by the {\it energetics argument}
put forward by Ramaty et al. (1997): if SN are the main source of GCR energy,
there is a limit to the amount of light elements produced per SN, which depends
on GCR and ISM composition, but also on the energy imparted to the GCR particles, i.e.
on the magnitude of the term $F^{GCR}$ in Eq. \ref{eqsimple}.  If the CNO content of {\it both} ISM and GCR becomes too low, there is simply not enough energy
in GCR to keep the Be yields constant.

\begin{figure}
\begin{center}
\includegraphics[angle=-90,width=0.49\textwidth]{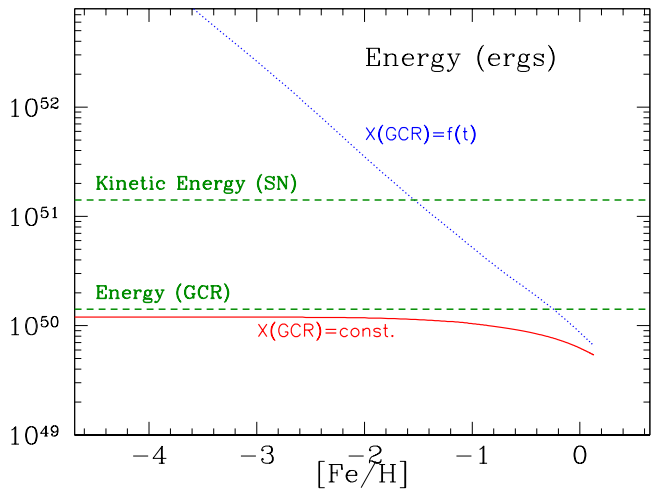}
\caption[]{Energy input required from energetic particles accelerated by one CCSN 
to produce a given mass of Be, such as
to have [Be/Fe]=0 (solar), assuming that a core-collapse SN produces, on average, 0.1 M$_{\odot}$
of Fe. {\it Solid} curve corresponds to the case of a constant composition for GCR, {\it dotted} curve corresponds
to a time-variable composition, following that of the ISM. In the former case, the required energy
is approximately equal to the energy imparted to energetic particles by supernovae, namely $\sim$0.1
of their kinetic energy of  $\sim$1.5 10$^{51}$ ergs; in the latter case, the energy required to
keep [Be/Fe]=0 becomes much higher than the total kinetic energy of a CCSN
for metallicities [Fe/H]$\leq$-1.6.}
\label{Fig:Be-energetics}
\end{center}
\end{figure}

Anticipating on the content of Sec. 3.3, we display in Fig. \ref{Fig:Be-energetics} 
the results of such a calculation. If the CNO content of GCR is assumed to be always constant
and equal to its present-day value, 
it takes 10\% of the SN kinetic energy, i.e. $\sim$10$^{50}$ ergs per SN, 
 to produce a constant yield of Be $y_{Be}\sim$10$^{-7}$ \ms; 
combined with a typical Fe yield of CCSN $y_{Fe}\sim$0.1 \ms, this  
leads to a production ratio [Be/Fe]=0, as observed.
In contrast, if the CNO content of GCR is assumed to decrease at low metallicities, following
that of the ISM, then below {\rm [Fe/H]=-1.6} it takes more than the total kinetic energy of a CCSN
 to obtain $y_{Be}\sim$10$^{-7}$ \ms. 
The only possibility left 
to achieve roughly constant LiBeB yields is then to assume that the ``reverse" component is primary,
i.e. that GCR have a roughly constant metallicity. This has profound implications for our 
understanding of the GCR origin, as we discuss in the next section. 
Of course,  before the aforementioned
Be and B observations,
no one would have had the idea to ask about  the GCR composition in the early Galaxy.

\subsection{The origin  of Galactic cosmic rays}

\begin{figure}
\begin{center}
\includegraphics[width=0.49\textwidth]{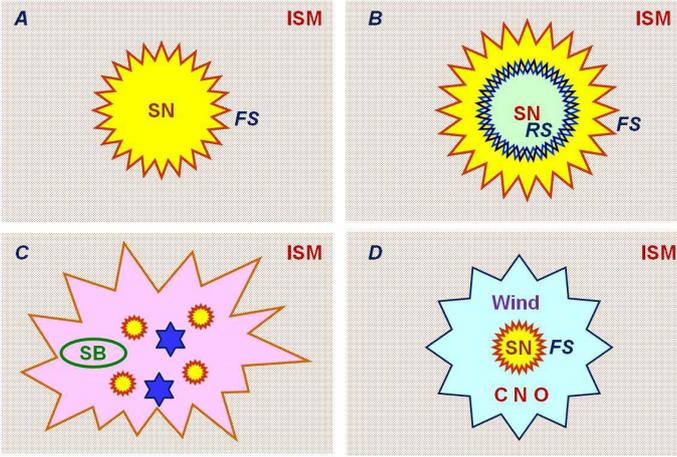}
\caption[]{ Scenarios for the origin of GCR. 
{\bf {\it A}}:  GCR originate from the interstellar medium (ISM) and are accelerated from the
forward shock (FS) of SN.
{\bf {\it B}}:  GCR originate from the interior of supernovae and are accelerated by the 
reverse shock (RS), propagating inwards.
{\bf {\it C}}:  GCR originate from superbubble (SB) material, enriched by the metals 
ejected by supernovae and massive star winds; they are accelerated by the forward shocks of 
supernovae {\it and} stellar winds.
{\bf {\it D}}:  GCR originate from the wind material of massive {\it rotating}
stars, {\it always rich in CNO} (but
not in heavier nuclei) and they are accelerated by the forward shock of the SN explosion.
}
\label{Fig:GCR-origin}
\end{center}
\end{figure}

For quite some time it was thought that GCR originate from the average ISM, where they
are accelerated by the {\it forward shocks} of SN explosions (Fig. \ref{Fig:GCR-origin}A). However, this can only produce secondary Be.

A  roughly constant abundance of C and O in GCR  can ``naturally" be understood if SN
accelerate their own ejecta through their {\it reverse schock} (Ramaty et al. 1997, see Fig. \ref{Fig:GCR-origin}B).  
However, the absence of unstable $^{59}$Ni 
(decaying through e$^-$ capture
within 10$^5$ yr) from observed GCR suggests that acceleration occurs 
$>$10$^5$ yr after the explosion (Wiedenbeck et al. 1999) 
when SN ejecta are presumably  already diluted in the ISM.  Furthermore, the reverse shock
has only a small fraction of the SN kinetic energy, while observed GCR require a large fraction
of it\footnote{The power of GCR is estimated to $\sim$10$^{41}$ erg s$^{-1}$ galaxywide, i.e.
about 10\% of the kinetic energy of SN, which is $\sim$10$^{42}$ erg s$^{-1}$ (assuming
3 SN/century for the Milky Way, each one endowed with an average kinetic energy of 1.5 10$^{51}$ ergs).}.

Taking up an idea of Kafatos et al. (1981), Higdon et al. (1998) 
suggested  that GCR are accelerated out of  {\it superbubble} (SB) material  
(Fig. \ref{Fig:GCR-origin}C), enriched by the ejecta of many SN as to have a large 
and  approximately constant metallicity. In this scenario, it is the forward shocks of SN 
that accelerate material ejected from other, previously exploded SN. Furthermore, it has been
argued that in such an environment GCR could be accelerated to higher energies 
 than in a single SN remnant (Parizot et al. 2004).  
That scenario has also been  invoked  to explain the present-day source isotopic 
composition of GCR (Higdon and Lingenfelter 2003; Binns et al. 2005, 2008). 
Notice that the main feature of that composition, namely
a high $^{22}$Ne/$^{20}$Ne ratio, is explained as being caused by the contribution of winds from Wolf-Rayet (WR) stars
(Cass\'e and Paul 1982), and the SB scenario apparently offers a plausible framework for bringing together
contributions from both  SN and WR stars.

 In a recent work, Prantzos (2012, hereafter P12) showed both qualitatively - on the basis of a simple nucleosynthesis
argument - and quantitatively  that superbubbles cannot be the main source of GCR acceleration.
Indeed, the main feature of the GCR source composition, namely the high \neo \ ratio, cannot be
obtained in a superbubble environment: the reason is that massive stars are the only source of both
$^{22}$Ne and $^{20}$Ne in the Universe and a full mixture of their ejecta - such as the one
presumably obtained in a superbubble - is expected to have a solar \neo \ ratio. This powerful qualititative
argument was substantiated in P12 by a detailed calculation of the evolving \neo \ ratio
in a superbubble, progressively enriched by the winds and the explosions of massive stars: only under 
unrealistically favourable circumstances and for a short early period is the \neo \ ratio in a superbubble
comparable to the observed one in GCR sources.

In that same work, P12 showed quantitatively how the GCR source ratio of \neo \ can be
explained by assuming acceleration by the forward shocks of supernova explosions, running through
the winds of massive stars and the ISM (Fig. \ref{Fig:GCR-origin}D). 
As already discussed, the idea of GCR \neo \ being due to
WR winds has been suggested long ago (Cass\'e and Paul 1982); however, the appropriate mixture
of wind and ISM material was always obtained by hand\footnote{Meyer (1985)
require a mixture of 1 part of He-burning material with 49 parts of normal ISM composition, Cass\'e and Paul (1982) and
Prantzos et al. (1987) infer 1 part of WC-star material with 50 parts of ISM,
while Binns et al.  (2005) require
a mixture of 80\% ISM with 20\% WR star material.}
while the connexion between the composition and the acceleration of GCR was left unclear.
P12 proposed a "unified" treatment, which brings together all massive stars (both low-mass ones
with negligible winds and massive ones with large mass losses, linked by the stellar
initial mass function) and which couples in a natural way   the
acceleration by the forward shock in the circumstellar medium  to the (time-dependent)
composition of accelerated particles: in the most massive stars, mostly wind material (rich in $^{22}$Ne)
is accelerated, while in the less massive stars mostly ISM (with solar $^{22}$Ne) is accelerated.
The main finding of P12 is that acceleration has to occur only in the early Sedov-Taylor phase of the
supernova remnant, for shock velocities higher than 1600-2000 km/s. 
Indeed,  to reproduce the GCR source \neo \ ratio, only a few tens of solar masses of circumstellar
material must be swept-up by the shock, otherwise the \neo \ ratio will be diluted to low values.

In this work we follow  the ideas of P12 and  
assume that during the evolution of the Galaxy,  GCR are accelerated mainly
by the forward shocks of supernovae, as they sweep up the massive star winds and the
ISM. The novel - and most important - feature of this work is that we
{\it calculate the evolving composition of GCR}
 (instead of assuming it to be constant, as previous workers in the field) by adopting  realistic models of 
rotating massive stars with mass loss, as we discuss in 
the next section.

\subsection{Wind composition of rotating massive stars}

The properties of rotating, massive stars are nicely summarised in the recent review paper
of Maeder and Meynet (2011) and are presented in considerable depth in the monograph of Maeder (2008).  
In particular, rotation has a twofold effect on stellar yields: it increases the size of the nuclearly processed layers
(since it mixes material further than convection alone) and reduces the escape velocity in the stellar equator,
allowing larger amounts of mass to be ejected into the wind. Both effects enhance the 
wind yields up to some mass limit; above it, the wind has removed so much mass
 that less material is left in the star to be processed in subsequent stages of the evolution,
  thus reducing the corresponding yields with respect to non-rotating models.

\begin{figure}
\begin{center}
\includegraphics[width=0.49\textwidth]{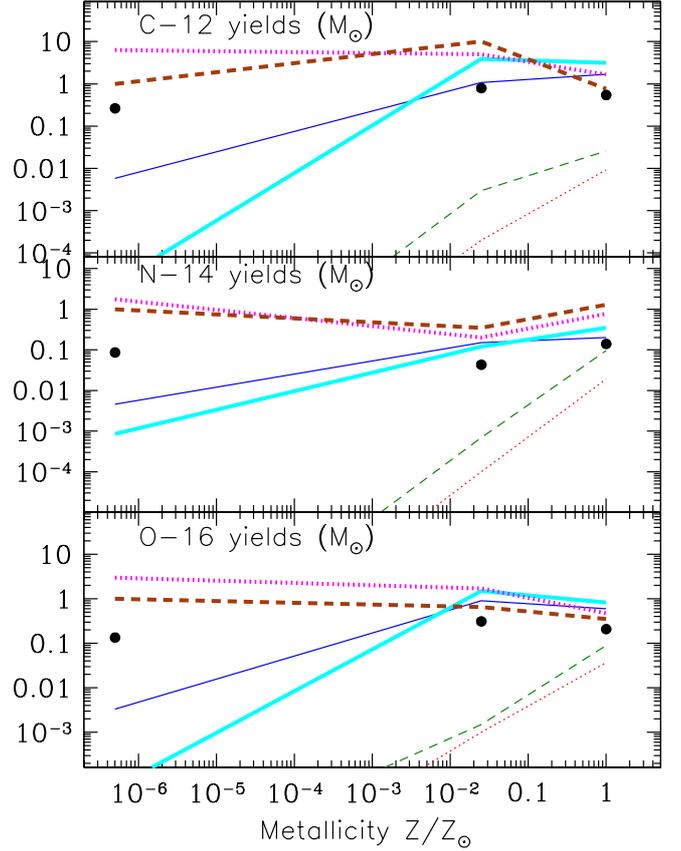}
\caption[]{ CNO content of the winds of rotating massive stars from the Geneva group.
In all panels, yields are for stars of 120 \ms ({\it thick dotted}), 80 \ms ({\it thick dashed}),
60 \ms ({\it thick solid}), 40 \ms ({\it thin solid}), 25 \ms ({\it thin dashed}) and 15 \ms
({\it thin dotted}). The curves connect yields that are provided at three values of the metallicity $Z$ 
($Z$=10$^{-8}$, 10$^{-5}$,  0.02). 
The {\it thick dots} correspond to yields averaged over a Salpeter IMF; average values
depend little on metallicity, because they 
are always dominated by the extremely high yields of the most massive stars.
}
\label{Fig:Wind-comp}
\end{center}
\end{figure}

To calculate the \neo \ ratio in present-day GCR, 
P12 adopted the models of the Geneva group 
(Hirschi et al. 2005), calculated for
solar metallicity. The initial rotational velocity of those models is
$v_{Rot}$=300 km/s on the ZAMS, corresponding to an average velocity of 220 km/s on the main sequence, 
i.e. close to the average observed value. For the purposes of this work, we
adopted a set of yields from the same group, extending down to a metallicity of $Z$=10$^{-8}$, i.e. $\sim$5 10$^{-6}$ \zs \
(Fig. \ref{Fig:Wind-comp}). In principle, this latter set is not quite
homogeneous, since  for the two lowest metallicities 
($Z$=10$^{-8}$, calculated in Hirschi 2006, and $Z$=10$^{-5}$, calculated in Decressin et al. 2007)
initial rotational velocities are $\upsilon_{init}$=800 km/s. As convincingly argued in Hirschi (2006), higher rotational velocities at
lower metallicities are a consequence of angular momentum conservation, since
less metallic stars are hotter and more compact than their higher metallicity
counterparts. This theoretical argument seems to be supported by observations
of rotating Be stars in the Magellanic Clouds and the Milky Way (Martayan et al. 2007),
although the observational situation is far from settled yet (see Penny and Gies 2009).

An indirect support for considerably higher rotational velocities at low metallicities
is provided by the observed evolution of nitrogen. The puzzle of the observed primary
behaviour of N vs Fe was known for a long time. Although intermediate-mass stars were known
to be able to provide primary N through hot-bottom burning (e.g. Renzini and Voli 1981),
these stars appear relatively late in the evolution of the Galaxy and cannot account
for the observations, especially after the VLT data of Spite et al. (2005) 
became available down to metallicities [Fe/H]$\sim$--3. 
In an early attempt using yields of rotating massive stars, Prantzos (2003a)
noticed that the then available N yields of the Geneva group, which concerned only
rotational velocities of 300 km/s across the full metallicity range, are too low
to reproduce the observations. In contrast, subsequently calculated yields 
at $Z$=10$^{-8}$ with velocities of 800 km/s (Hirschi 2006) produce
much more N at low $Z$ and  are indeed able to
reproduce the observed evolution of N, as shown in Chiappini et al. (2007).
This result should by no means be considered as a proof of the validity 
of the concept of these high rotational velocities at low $Z$, but it certainly
constitutes an encouraging hint towards that direction and we adopt those same yields
in this work.

The main feature of Fig. \ref{Fig:Wind-comp} is that the yields of the rotating massive stars, when 
averaged over a stellar initial mass function (the one of Salpeter being adopted here)
show a remarkable constancy with metallicity: the winds of those stars expel about
the same overall amount of C, N and O nuclei at all metallicities. If GCR are
accelerated from such material, then they can naturally provide primary Be, as observed.
Notice, however, that the actual calculation of the GCR composition from the aforementioned
yields is not straightforward; we present it in some detail in Sec. 3.2.

\section{Model and ingredients}
\label{Sec:EvolBeB}

\subsection{Model of galactic chemical evolution}
\label{Subsec:Model}

The model adopted here is an updated version  of  the one presented in
Goswami and Prantzos (2000, hereafter GP2000). 
The set of chemical evolution equations is solved without the instantaneous recycling
approximation (IRA), for two galactic systems representing the halo and the local disc,
respectively. The halo is formed on a timescale of 1 Gyr with a star formation rate
(SFR) proportional to the gas mass and an outflow rate equal to eight times the SFR; the latter
ingredient is necessary in order to reproduce the observed halo metallicity distribution
(see Prantzos 2003b and references therein). 
The local disc is formed with the same prescription as for the SFR, but on a longer timescale of 8 Gyr,
allowing one to reproduce the corresponding metallicity distribution; the combination of
SFR and infall rates reproduces the present-day local gas fraction of $\sigma_{Gas}\sim$0.2.

  \begin{figure}
   \centering
 \includegraphics[width=0.49\textwidth]{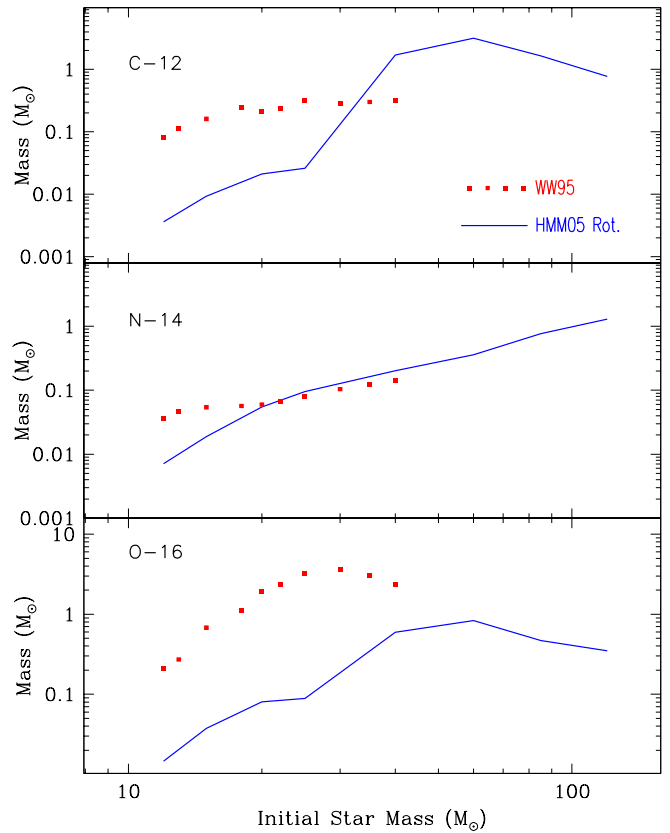}
 \caption{CNO yields of massive stars of solar initial metallicity adopted in this work: total yields 
 for stars with no mass loss and no rotation in the 11-40 \ms \ range, from WW95 ({\it points});
  and wind yields of rotating mass-losing stars in the 12-120 \ms \ range, 
  from Hirschi et al. (2005, {\it curves}). 
 }
         \label{Fig:CNO_yields}
   \end{figure}

  \begin{figure}
   \centering
 \includegraphics[width=0.49\textwidth]{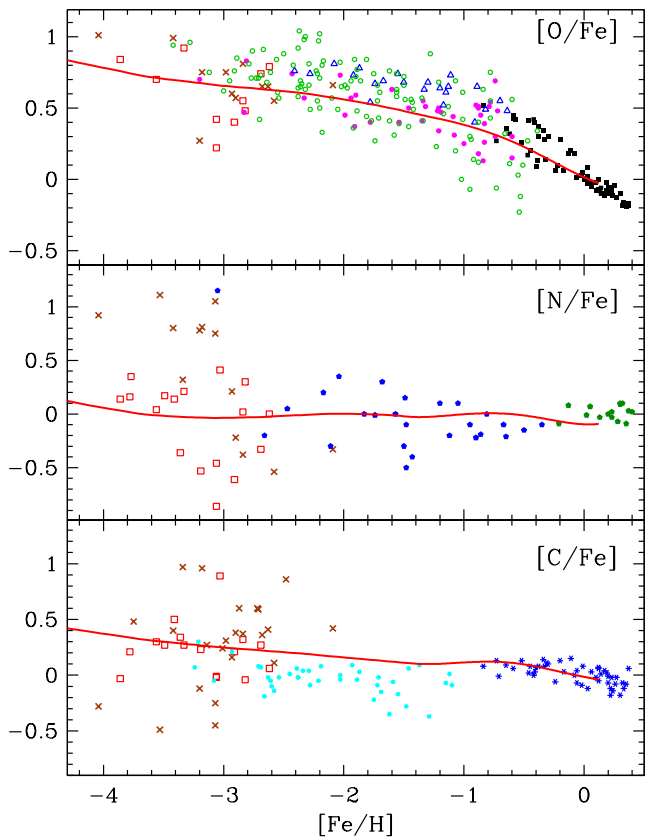}
 \caption{Evolution of O/Fe, N/Fe and C/Fe vs Fe/H.
 Data sources:  Spite et al. (2005, {\it open squares} for C,N,O), Lai et al. (2008,  X for C,N,O), 
 Primas (2010, {\it filled circles} for O), Tan et al.  (2009, {\it open triangles} for O), Boesgaard et al.  (2011, {\it open circles} for O), Bensby et al.
 2003, {\it filled squares} for O), Israelian et al. (2004, {\it filled pentagons} for N), 
 Bensby et al. (2006, {\it asterisks} for C),
 Fabbian et al.  (2009, {\it filled circles} for C).
 }
         \label{Fig:CNO_ChemEvol}
   \end{figure}

The IMF is taken from Kroupa (2003) and extends from 0.1 \ms \ to 120 \ms.
The adopted yields are from van den Hoek and Gronewegen (1997) for low and intermediate mass 
stars and taken from Woosley and Weaver (1995, hereafter WW95) for stars in the 11-40 \ms \ range, where stellar winds
play a negligible role even at solar metallicity. For stars more massive than 40 \ms, instead of extrapolating the WW95
yields, we adopted the yields of rotating, mass-losing stars of the Geneva group for
 H, He, C, N, and O, as described in Sec. 2.4. The adopted massive star yields at solar metallicity appear
 in Fig.  \ref{Fig:CNO_yields}. It can be seen that the contribution of the winds to the O yields is negligible, but
 it becomes substantial for C and N; it becomes dominant for N at low metallicities, because N is produced
 as secondary in WW95. 
 All the yields are metallicity-dependent and they are properly interpolated in mass and metallicity. 
This concerns also the yields of $^7$Li and $^{11}$B from massive stars, which are produced by $\nu$-induced
nucleosynthesis in WW95 and will be further discussed in Sec. 4.2 and 5.1, respectively. It is assumed that the
ejecta of stars of all masses contain no \lib, Be or $^{10}$B, i.e. that those fragile isotopes are
astrated with a 100\% efficiency. Those nuclides suffer more from astration  than deuterium - which receives a
continuous contribution from infalling gas of primordial composition - and their astration has to be included
in chemical evolution models. Low-mass stars may be net producers  of \lia, at least
within some mass range (see discussion in Sec. 5.1).
Yields for SNIa are taken from Iwamoto et al. (1999), while the SNIa rate 
(important for the evolution of Fe)  follows the prescription of Greggio (2005) for the single-degenerate scenario.

As described in GP2000, the model reproduces the main features
of the local halo and disc well, including the absolute abundances of most elements and isotopes between C
and Zn at the Sun's formation 4.5 Gyr ago. In Fig. \ref{Fig:CNO_ChemEvol} it can be seen that
it reproduces quite satisfactorily the full evolution of the abundance
of all three elements that contribute to LiBeB production, namely C, N and O.
In the case of N, the success is due to the adoption of the Geneva yields of rapidly
rotating stars at low metallicities (unavailable at the time of GP2000), as already
shown by Chiappini et al. (2007). Fig. \ref{Fig:CNO_ChemEvol} indicates then that the evolution
of the {\it direct} component of the LiBeB production will be consistently calculated, since the
ISM abundances of all CNO elements as a function of metallicity agree with the observations.
The only other work on LiBeB presenting the ISM abundances of all three CNO
elements is the one of Alibes et al. (2002), where early N was calculated as secondary.

\subsection{Composition of GCR }
\label{Subsec:LiBeBProd}

It was realised early on that the {\it elemental} composition of GCR {\it at the source} (i.e. after accounting
for propagation effects) differs significantly
from that of the ISM. 
Volatiles behave  differently from refractories: the former display a mass-dependent enrichment
with respect to H, which reaches a factor of 10 for the heaviest of them; the latter are all
overabundant (w.r.t. H) by a factor of 20, while C and O display intermediate overabundances,
by factors of 9 and 5, respectively (e.g. Wiedenbeck et al. 2007 and references therein). Finally, He is slightly
underabundant, with (He/H)$_{GCR}$/(He/H)$_{\odot}$=0.8 

This complex pattern is now thought to result not from ionisation effects (as suggested in Cass\'e and
Goret 1978, and further developped by Meyer 1985) but rather from effects related to elemental
condensation temperature (Meyer et al. 1997). Supernova shocks pick up and accelerate gas ions and dust grains
simultaneously. The gas ions are accelerated directly to
cosmic-ray energies in the shock, which produces
an enhancement of ions with higher mass/charge ratios (i.e., heavier elements).
On the other hand, refractories are locked in dust grains, which are
sputtered by repeated SN shocks and the released ions are easily picked-up and accelerated (see Ellison et al. 1997
for a detailed presentation of the model).
This quite elaborate scheme, which builds on earlier ideas by e.g.  Bibring and Cesarsky (1981),
accounts quantitatively for most of the observed features of GCR source composition.

  \begin{figure}
   \centering
 \includegraphics[width=0.49\textwidth]{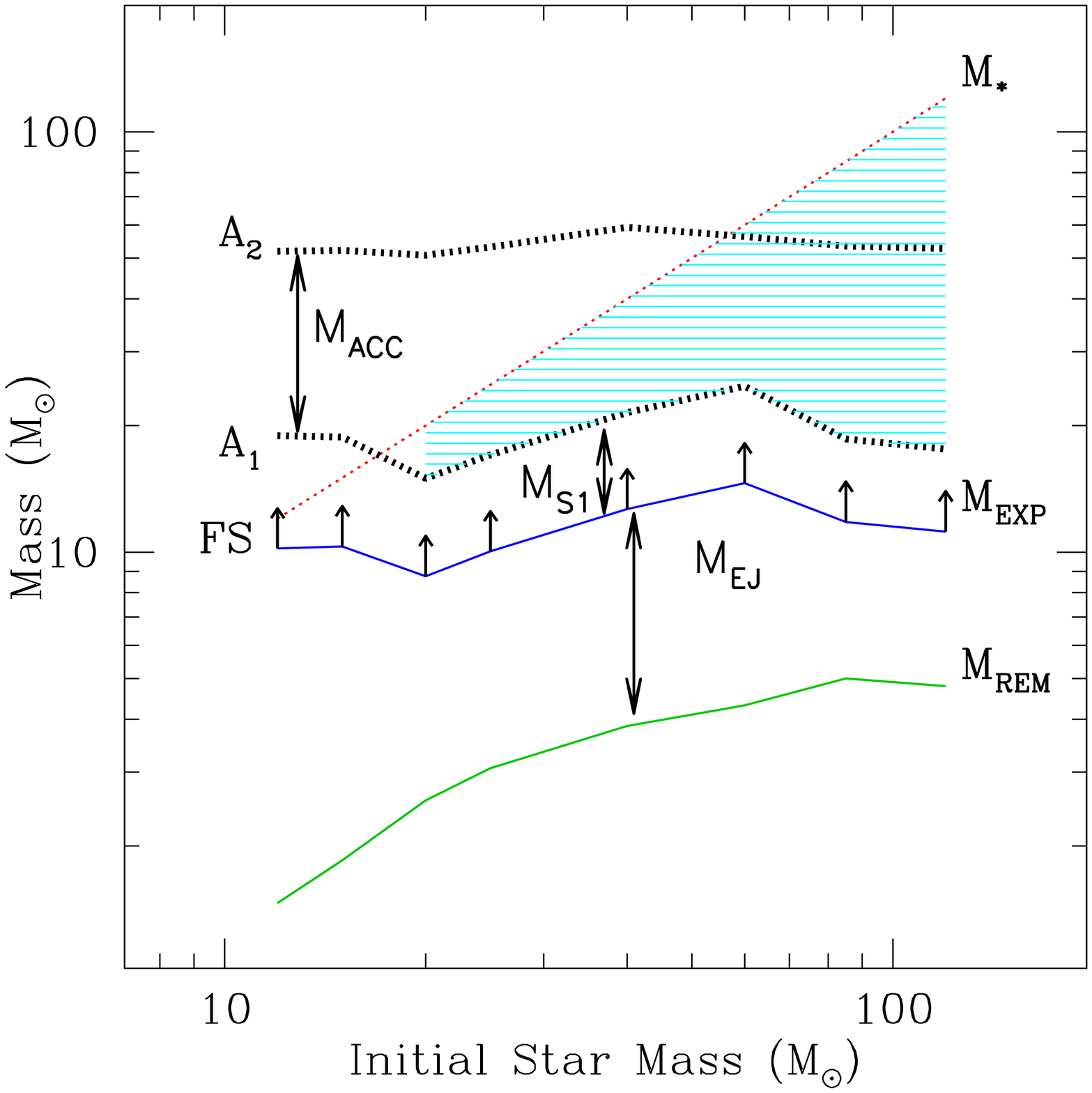}
 \caption{Illustration of the scheme adopted for the acceleration of GCR in Prantzos (2012) and here. 
 At explosion, the star of initial mass $M_*$ has a mass $M_{EXP}$ and leaves behind
 a remnant of mass $M_{REM}$, i..e the explosively ejected mass is $M_{EJ}$, while the mass
 previously ejected  by the wind is $M_{Wind}$=$M_*$-$M_{EXP}$. 
 Particle acceleration
 starts at the beginning of the Sedov-Taylor phase, located at mass coordinate $A_1$=\mex + \mej, i.e. when
 the forward shock (FS, {\it arrows}), launched at \mex,  has swept up a mass $M_{S1}$=\mej. Acceleration stops
 at mass coordinate $A_2$, selected in P12  to reproduce - after an 
  average with a Salpeter IMF - the GCR source
 ratio of $^{22}$Ne/$^{20}$Ne=5.3  in solar units;  
 in the case of rotating stars adopted  here, it corresponds  to a shock velocity of 1900 km/s.
 The mass sampled by the FS between those two regions is $M_{Acc}$=$A_2$-$A_1$. For rotating stars with
 mass $M>$30 \ms, an increasing part of $M_{Acc}$ includes nuclearly processed material ({\it shaded aerea}),
 while for rotating stars with $M<$18 \ms, $M_{Acc}$ contains only material of ISM composition.
 The yields of Fig. 4 correspond to the mass $M_{Wind}$=$M_*$-$M_{EXP}$, while the composition of GCR
 adopted here corresponds to the mass $M_{ACC}$.
 }
  \label{Fig:AccelvsM}
   \end{figure}

The most conspicuous feature of GCR source composition   is undoubtely the high isotopic \neo \ ratio,
which is 5.3$\pm$0.3 times the value of the (\neo)$_{\odot}$ \ ratio in the solar wind (Binns et al. 2008).
Contrary to the case of the elemental source GCR abundances, which may be affected by various
physico-chemical factors (first ionisation potential, condensation temperature, etc.),
isotopic ratios can only be affected by nucleosynthetic processes and thus provide crucial
information on the origin of cosmic ray particles. P12 showed how this ratio  can be explained
quantitatively by assuming that the forward shocks of supernovae accelerate circumstellar material of mass-losing stars,
which is composed either of nuclearly processed material (rich in \neb, in the case of stars with mass M$>$40 \ms)
or by pure ISM (with normal \neb, in the case of stars with M$<$30 \ms). To obtain the observed GCR source \neo \
ratio, P12 found that acceleration had to occur only in the early Sedov-Taylor  phase of the supernova
for shock velocities higher than $\sim$1900 km/s for the rotating stars of the Geneva group\footnote{If acceleration proceeds
at lower shock velocities,  too much ISM material is accelerated and the resulting \neo \ ratio becomes too low.}. This value
is obtained after {\it all} stars between 10 and 120 \ms \ are considered and the corresponding composition is averaged over the
IMF. 

Fig. \ref{Fig:AccelvsM} illustrates the scheme adopted in P12. 
In that work it was shown that this scheme, although nicely fitting the
GCR \neo \ ratio,  cannot directly reproduce the observed GCR abundancs of CNO elements. The resulting C/O and N/O ratios
(expressed in solar units) are higher than the corresponding GCR source ratios by factors of $\sim$2 (see Fig. 6 in P12).
The reason is that the winds of massive stars are loaded essentially with products of H-burning (He and N) and early He-burning
(C and \neb), while O is a product of late He-burning and appears only marginally in the stellar winds. P12 argued that
this apparent disagreement can be easily understood in the framework proposed by Meyer et al. (1997) and Ellison et al. (1997):
the  (mildly) refractory O is more easily picked up and accelerated by the shock than the less refractory C and the volatile N.
Its abundance in GCR is then enhanced and the calculated C/O and N/O ratios (shaped by nucleosynthesis and mass loss) 
have to decrease accordingly, modulated by atomic effects of shock acceleration.

\begin{figure}
   \centering
 \includegraphics[width=0.49\textwidth]{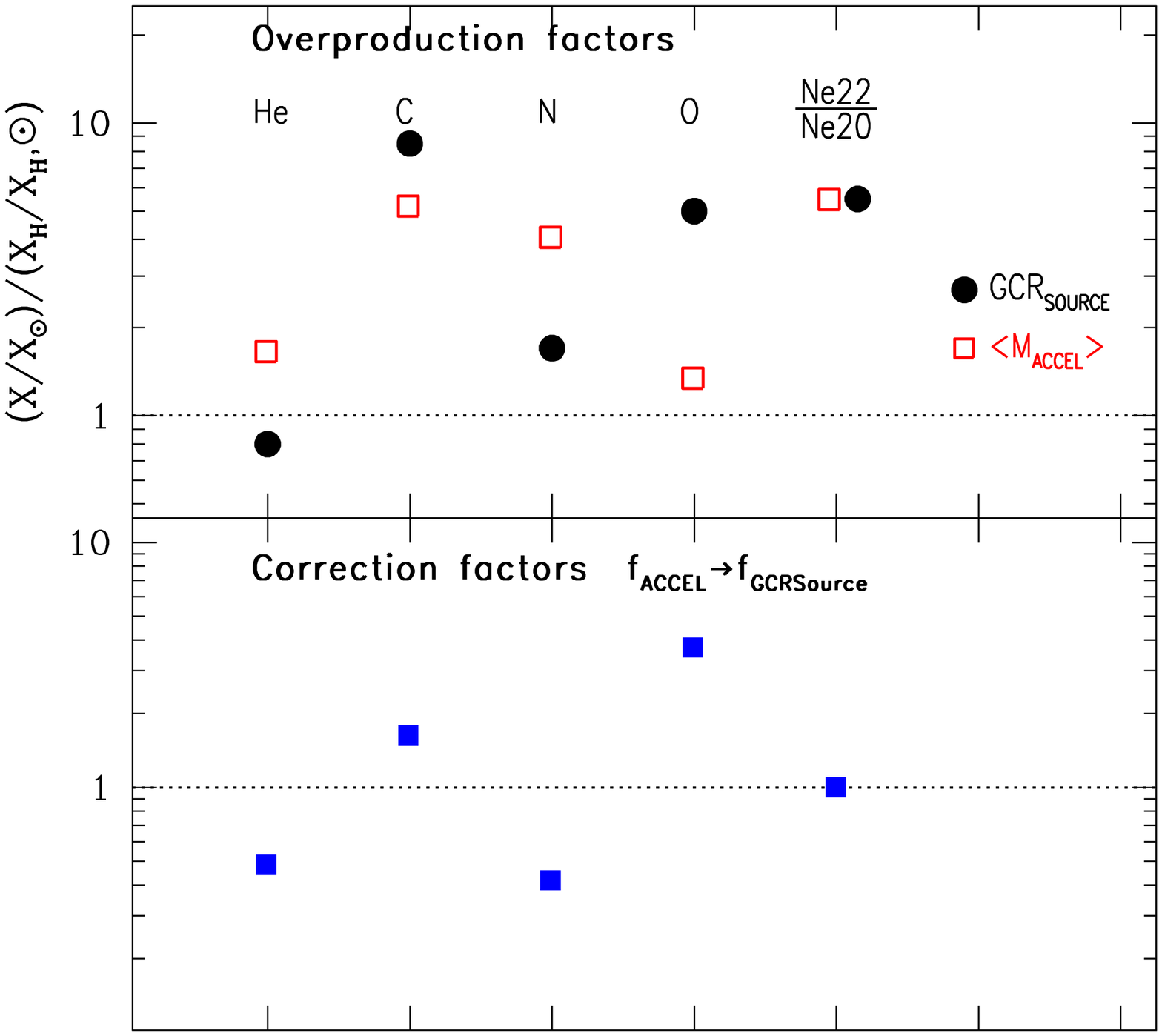}
 \caption{Source abundances in GCR ({\it filled circles}) and in the
  particles accelerated in the mass $M_{Acc}$ of Fig. 6 ({\it open squares}).
  They are normalised to the abundance of hydrogen and expressed in the
  corresponding solar units: $f=\frac{X/X_{\odot}}{X_H/X_{H,\odot}}$. 
  By construction, the $^{22}$Ne/$^{20}$Ne
  ratio - unaffected by atomic affects - matches the GCR source value.
  For He, C, N and O, a correction factor (simulating atomic effects)
  $R_{cor}=f_{GCR}/f_{M_{ACC}}$ is calculated and applied to the accelerated 
  composition of $M_{ACC}$ at all metallicities (see bottom panel of Fig. \ref{Fig:GCR_CNOcomp}). 
 }
         \label{Fig:Accel-eff}
   \end{figure}

This modulation appears in Fig. \ref{Fig:Accel-eff}. The abundances of He, C, N and O (with respect to H and in solar units) appear
in the upper panel for both the GCR sources and the accelerated material, according to 
the model results of P12, which are obtained
for the Geneva models  at solar metallicity. Clearly, O/H is much higher in GCR sources than in
accelerated material and C/H only slightly so, while for volatiles the ratios  He/H and N/H are lower in GCR sources than in the model.

The lower panel of Fig. \ref{Fig:Accel-eff} displays the correction factors $R_{cor}$ required to bring into agreement the model results ($f_{M_{ACC}}$)
with the GCR source abundances ($f_{GCR}$): a large enhancement ($R_{cor}$=4) has to be aplied to O (which is efficiently accelerated as a refractory),
a smaller enhancement to the less refractory C and depletion factors ($R_{cor}<$1) to both volatiles He and N. Those corrections, which
are justified from the current paradigm of GCR acceleration, bring into agreement the model results with the observed
GCR source values of {\it all} elements participating in LiBeB production through the {\it reverse} component.
% namely $Y^{GCR}_H/Y$=1, $Y^{GCR}_{He}$=He/H=,  $Y^{GCR}_{C}$=C/H=, $Y^{GCR}_{N}$=N/H= and $Y^{GCR}_{O}$=O/H= NA TA VALW SWSTA....

Such a GCR source composition, valid for the present-day cosmic rays, has been adopted in some previous works in the subject (e.g. Ramaty et al. 1997\footnote{Ramaty et al. (1997) adopted a GCR source
composition from Engelman et al. (1990), slightly more enhanced in C and O than in this work, 
where the composition of Meyer et al. (1997) is adopted.}).
In those works it was {\it assumed}  that at lower metallicities {\it the same GCR composition applies also} (as suggested by the
observed linearity of Be vs. Fe) and a theoretical justification for that was invoked, namely the superbubble scenario for GCR
acceleration. In this work we dispense with these assumptions, but we actually {\it calculate} the GCR source composition as a function
of metallicity, by using the wind composition of pre-supernova stars
(provided by the Geneva rotating models of Sec. 2.4) and the ISM composition (provided
by our model of Sec. 3.1): in each time step we calculate the composition of accelerated particles (as
a mixture of ISM and wind compositions, properly weighted by the IMF according to the
scheme of Fig. \ref{Fig:AccelvsM}) and we apply to it the correction factors of Fig. \ref{Fig:Accel-eff}. 

\begin{figure}
\begin{center}
\includegraphics[width=0.49\textwidth]{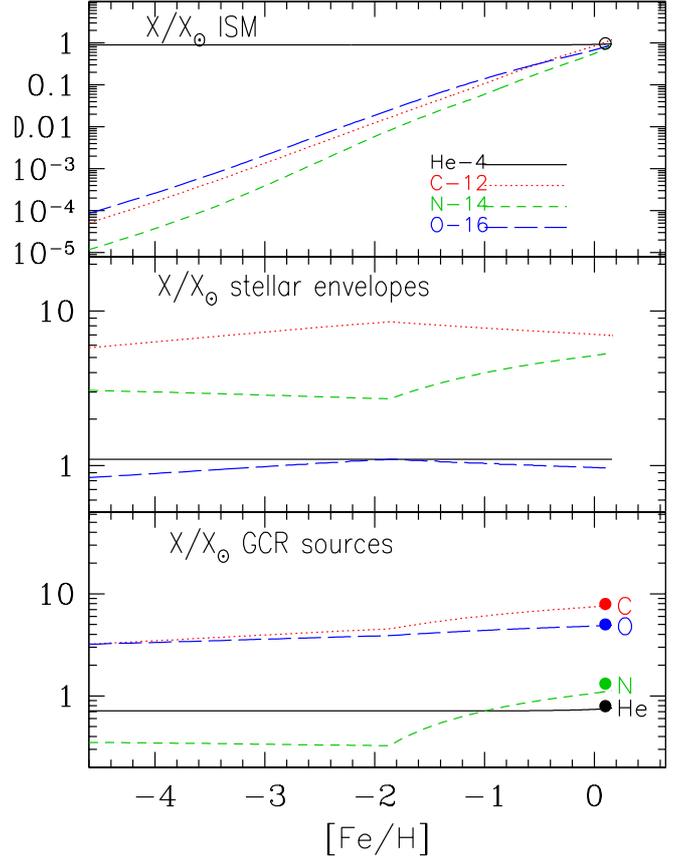}
\caption[]{ Evolution of the chemical composition (normalised to  corresponding 
solar abundances) of He ({\it solid}), C ({\it dotted}), N ({\it short dashed})
and O ({\it long dashed})in: ISM ({\it top}), massive star winds ({\it middle}) and
GCR ({\it bottom}); the latter is the one calculated for the accelerated particles of
$M_{ACC}$ (Fig. 6) corrected to reproduce GCR source abundances today (as in Fig. 7).
  {\it Dots} in lower panel indicate estimated GCR source composition {\it today},
from Ellison et al. (1997). 
}
\label{Fig:GCR_CNOcomp}
\end{center}
\end{figure}

The results appear in Fig. \ref{Fig:GCR_CNOcomp}, displaying the He, C, N, and O abundances of the
evolving ISM (top), of the stellar winds (middle) and of the GCR sources (bottom). The latter fits well
 the observed GCR source composition today {\it by construction} (i.e. through the application of the
correction factors discussed in the beginning of this subsection).

After calculating in a consistent way the abundances of all key elements that produce LiBeB both in the ISM and in GCR,
we proceed with the calculation of spallogenic LiBeB production in the next section.

\subsection{Production of LibeB by GCR }
\label{Subsec:LiBeBProd}

The abundances by number $Y^{ISM}_L=N_L/N_H$ (per hydrogen atom of the ISM) of the light nuclides $L$
during the evolution of the Galaxy are calculated by
\begin{equation}
\frac{\vartheta Y_L}{\vartheta t} \ = \sum_j Y_j^{ISM} \ \sum_i \int_T^{\infty} F_i^{GCR}(E)
\ \sigma_{ij}^L(E) \ P_{ij}^L(E_P) \ dE.
\end{equation}
In this expression, $L$=1,...,5 for $^6$Li,$^7$Li, $^9$Be, $^{10}$B and $^{11}$B. 
The indices $i$ and $j$ run over the range 1,...,5 for H, $^4$He, $^{12}$C, 
$^{14}$N, and $^{16}$O. The omnidirectional {\it flux} of GCR particles 
\begin{equation}
F_i(E) \ = \ N_i(E) \ v(E)
\end{equation}
(where $v(E)$ is the particle velocity as a function of the {\it energy per nucleon E}), 
is obtained by assuming that the {\it propagated energy spectrum } $N_i(E)$
of GCR reaches equilibrium, i.e. that the various losses (through escape, ionisation, spallation etc.)
just balance continuous injection from sources with an {\it injection  spectrum} $Q_i(E)$; the 
equilibrium solution for {\it primary nuclei}, like H,He,C,N,O, as
obtained e.g. in MAR, is
\begin{equation}
N_i(E) \ = \ \frac{1}{b_i(E)} \ \int_E^{\infty} Q_i(E') \ {\rm exp} 
\left[-\frac{R_i(E')-R_i(E)}{\Lambda} \right] \ dE' ,
\end{equation}
where 
\begin{equation}
R_i(E)=\int_0^E \frac{\rho \ v(E')}{b_i(E')} \  dE'
\end{equation}
is the {\it ionisation range}, with $\rho$ the ISM mass density and
$\Lambda = \rho v \tau_{ESC}$ is the {\it escape length} from the Galaxy,
$\tau_{ESC}$ the corresponding time scale, and $b(E)$
represents ionisation losses. The functions $b(E)$ and $\tau(E)$ are determined from basic physics
and from the observed 
properties of the ISM (density, composition, ionisation stage)
and of the CR (abundance ratios of primary to secondary and of unstable
to stable nuclei), see e.g. Strong et al. (2007) and references therein.

\begin{figure}
\begin{center}
\includegraphics[width=0.49\textwidth]{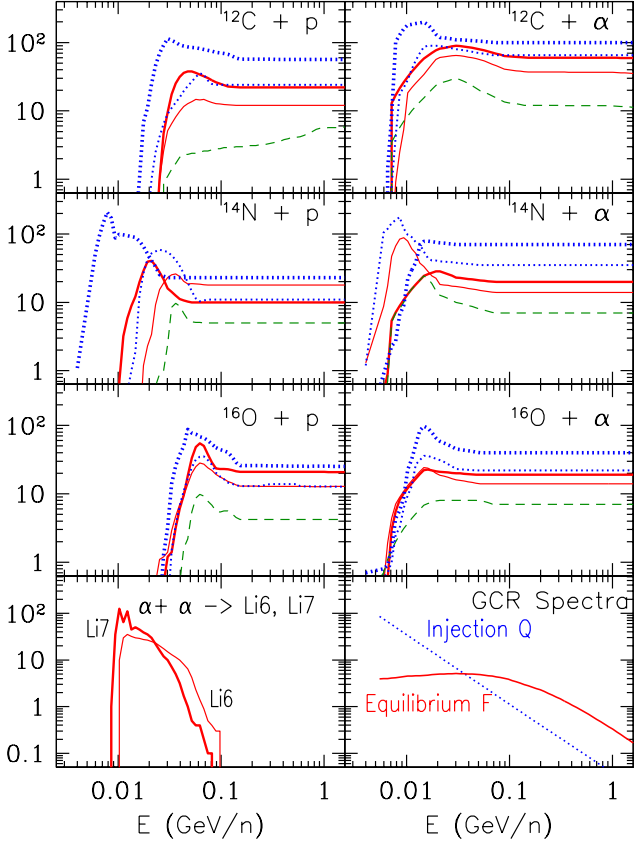}
\caption[]{ Cross sections (mb) for the production of Li, Be and B by spallation of
CNO nuclei with protons and alphas, as a function of particle energy per nucleon. 
Data are from Read and Viola (1984) and  Mercer et al (2001, $\alpha+\alpha$ reactions).
In all panels {\it thick dotted} curves correspond to production of $^{11}$B, {\it thin dotted}
to $^{10}$B, {\it thick solid} to $^7$Li, {\it thin solid} to $^6$Li and {\it dashed} to $^9$Be.
In the {\it bottom right} panel appear the adopted GCR spectra (arbitrary units): injection $Q$
({\it dotted}) and equilibrium $F$ ({\it solid}) .
}
\label{Fig:CrossSec}
\end{center}
\end{figure}

The cross sections
$\sigma_{ij}^L(E)$ represent the probability of producing nucleus $L$
through the interaction of nuclei $i$ and $j$, and they have a threshold $T$
(Fig.  \ref{Fig:CrossSec}).

The quantities
 $ P_{ij}^L(E_P)$ represent the fraction of light nuclei $L$ that are produced
at energy $E_P$ and are incorporated in the ISM. They are given by  
\begin{equation}
 P_{ij}^L(E_P) \ = \ {\rm exp}\left[-\frac{R_L(E_P)}{\Lambda}\right] 
\end{equation}
where $R_L(E)$ is the ionisation range of nucleus $L$. The energy 
$E_P$ is close to zero when a fast proton or alpha hits a CNO nucleus
of the ISM (i.e. the resulting light nucleus is created at rest and $P\sim$1),
and $E_P=E$ when fast CNO nuclei are spallated by ISM protons and alphas
(i.e. the resulting light nuclei inherit the same energy per nucleon).
For the fusion reaction $\alpha + \alpha \longrightarrow$ $^{6,7}$Li
($i=j=$2) the resulting Li nuclei are created with a velocity about half that
of the fast $\alpha$ particles, and $E_P=E/4$ (see Eq. (6) in MAR).

The CR equilibrium spectrum is known very poorly at low energies, precisely 
those  that are important for LiBeB production (in view of the relevant production
cross sections, see Fig. \ref{Fig:CrossSec}). The reason is the poorly understood
modulation effects of the solar wind. Instead of using a demodulated spectrum
(e.g. Ip and Axford 1985), in most studies of Li production, a theoretical
injection spectrum is adopted and propagated in the Galaxy  to recover
the equilibrium spectrum through Eq. (3). The form of the injection spectrum is
motivated by theories of collisionless shock acceleration 
(e.g. Ellison and Ramaty 1985) and we adopt here the frequently used 
(Prantzos et al. 1993; Fields et al. 1994, 2001;  Ramaty et al. 1997, 2000)
 momentum spectrum of the form

\begin{equation}
Q_i(E) \ = Y_i^{GCR} \ \frac{p^{-s}}{\beta} {\rm exp}(-E/E_C),
\end{equation}
where $\beta=v/c$ is the velocity expressed as a 
fraction of the light velocity,
$p$ the particle momentum per nucleon, 
the factor $s$ is usually 2$<s<$3 (in the case
of strong shocks) and we adopte $s$=2.3 here and $E_C$ is a cut-off energy that we take here
to be $E_C$=1 TeV. The resulting injection and equilibrium spectra (after propagation through a "canonical"
path length of $\Lambda$=10 gr cm$^{-2}$ appear in
 Fig.   \ref{Fig:CrossSec} (bottom right panel).

The total power (energy per unit time) in accelerated particles is
\begin{equation}
\dot{W} \ = \frac{\vartheta W}{\vartheta t}  \ 
= \ \sum_i A_i \ \int^{\infty}_0 \ E \ Q_i \ dE,
\end{equation}
where multiplication by the mass number $A_i$ accounts for the
fact that energy $E$ is always expressed in units
of energy/nucleon. This power is provided by the main
energy source of GCR, namely supernovae. Theoretical arguments
suggest that a fraction $e_{GCR}\sim$0.1 of the kinetic energy $E_K\sim$1.5 10$^{51}$ erg of supernovae
goes into GCR acceleration. Thus
\begin{equation}
\dot{W} \ =  \ e_{GCR} \ E_K \ SNR,
\end{equation}
where SNR is the rate of supernovae (number of SN explosions per unit time) given by
the model. 

The link between the chemical evolution model and the local production of LiBeB by GCR
is provided by Eqs. (2) to (7) concerning the chemistry ($Y_i^{ISM}$ appears in Eq. (2) and
$Y_i^{GCR}$ in Eq. 7) and through Eqs. 
(8) and (9) concerning the energetics (since the latter equations allow one to normalise
the injection spectra $Q(E)$). The self-consistent calculation of
the "coupling term"  $Y_i^{ISM}$ has firstly been performed 
in Prantzos et al. (1993) and that of the energetics in Ramaty et al. (1997)
but it is the first time that the term $Y_i^{GCR}(t)$ is {\it calculated}
on the basis of realistic models (see Sec. 3.2) and not just assumed, as in all previous studies.

Before proceeding to present the results of our model, we notice that the uncertainties due to the nuclear physics
are small. Indeed, the uncertainties in the adopted cross-sections
are, in general, smaller than 10\% (at least at low energies, where most of the GCR particles reside, see
discussion in Mercer et al. 2001). Moreover, Kneller et al. (2003)  investigated the key approximations made in the
simplified calculation (Eqs. 2 to 6) from the nuclear physics point of view and  found
that the  introduced  errors are negligible. Thus, the overall uncertainties of our calculation will be
 almost exclusively of astrophysical origin.

\section{Evolution of Be and B}
\label{sec:LiBeBProd}

This section presents the results for the evolution of Be and B of 
the chemical evolution model of Sec. 3.1, augmented with the calculation of GCR composition of Sec. 3.2
and the spallogenic production of LiBeB of Sec. 3.3.

\subsection{Be evolution}

\begin{figure}
\begin{center}
\includegraphics[width=0.49\textwidth]{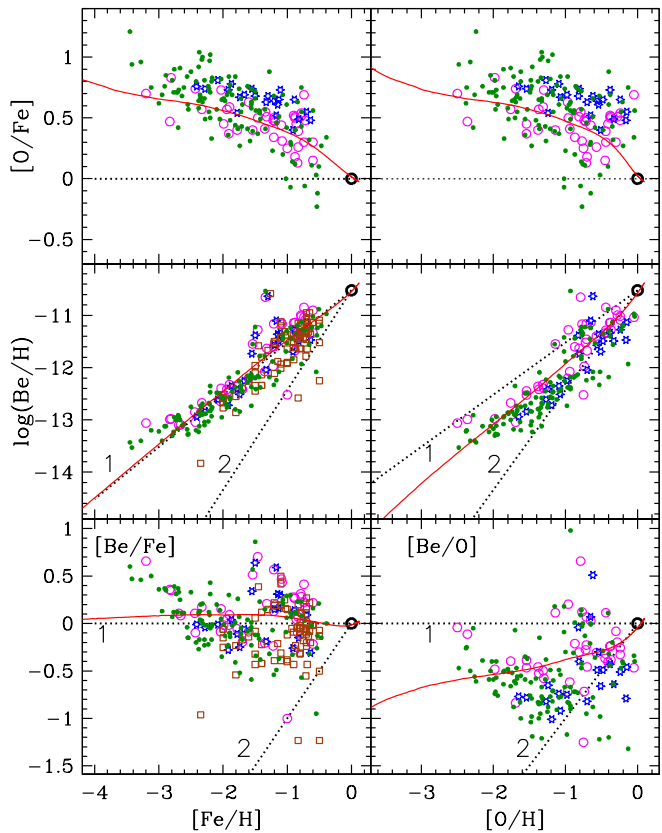}
\caption[]{Evolution of O/Fe ({\it top}), Be/H ({\it middle}) 
and Be/Fe bottom, as a function of Fe ({\it left}) and O ({\it right}).
{\it Dotted} lines in middle and bottom panels indicate primary and secondary
evolution, respectively. Data sources are as in Fig. 1, but only stars with $both$ O and Be detected are displayed here.
}
\label{Fig:Be-evol}
\end{center}
\end{figure}

The evolution of Be is presented in the middle and bottom panels of Fig. \ref{Fig:Be-evol}
as a function of Fe/H and O/H, respectively. Evidently, the adopted prescriptions
lead to a Be/H evolution that is linear with respect to Fe/H, to a very high accuracy:
over four decades in [Fe/H] (from -4 to 0) the differences from linearity are less than 0.1 dex,
as can be best seen in the bottom-left panel of Fig. \ref{Fig:Be-evol}.
This result is due to two factors:

i) The adopted GCR energetics and Fe yields, which are independent of the metallicity.
At all metallicities, it is assumed that core-collapse SN release $\sim$0.07 \ms \ of Fe (the average
of WW95 yields) and accelerate GCR with an average energy of 1.5 10$^{50}$ ergs.

ii) The combination of the compositions of the evolving ISM and of the winds of {\it rotating} massive stars
which render the resulting GCR composition approximately metallicity independent 
(notice that the slow increase of C and O in GCR - see  bottom panel of Fig. \ref{Fig:GCR_CNOcomp} 
- plays some role in the late behaviour of LiBeB isotopes, as discussed below).

Notice that, although ingredients (i) and (ii) play a key role in the resulting linearity of Be/H with Fe/H, the solution they provide
is - at least formally - degenerate: 
the same result would be obtained if e.g. Fe yields and GCR energies were allowed to vary in lockstep
during galactic evolution (i. e. the present-day values of those quantities could be reached after starting from
values 10 times higher or lower in the early Galaxy). In that case, the variation in the Be production would be compensated
by a similar variation in the Fe production, leaving the Be/Fe ratio intact\footnote{The same variation should be then assumed
for the yields of {\it all} other elements $X$, as to leave the $X/Fe$ ratio intact.}. A similar degeneracy would occur
if variation in GCR energetics were accompanied by a similar variation in GCR composition. In that case, 
 the equality of Eq. 8 would be preserved with  both members varying in the same way, i.e. by assuming a variation of $\dot{W}(t)$ (from  Eq. 9) equal to the variation of $Q_i(E,t)$ (from $Y_i^{GCR}(t)$ in Eq. 7). Although the combined variation
 of those properties cannot be excluded\footnote{One may well imagine that in the early Galaxy stars produced smaller amounts of wind yields (and thus smaller $Y_i^{GCR}$) and that stellar explosions were weaker (because
pre-supernova stars were heavier due to lower mass losses),  providing lower kinetic energies 
$E_K$ and  lower Fe yields.}, such a precise synchronisation 
appears improbable
and we shall consider here only our basic scenario, i.e. quasi-constant yields of primary elements and constant
SN energetics.

In Fig. \ref{Fig:Be-evol} we also present the evolution of Be vs. O and O vs. Fe, but in contrsast to the case of 
Fig. \ref{Fig:Be-obs} we display O vs. Fe observations only for the stars with detected abundances of Be. Clearly,  
Be correlates better with Fe than with O, both observationally and theoretically. This may appear puzzling at first sight, because
Be is produced by spallation of O, not  Fe.  However, O in the ISM (now seen in halo stars) is produced by metallicity-independent
yields of massive stars, while Be in ISM (seen in halo stars) is produced by slowly increasing O of GCR (bottom
panel in Fig. \ref{Fig:GCR_CNOcomp}). The amount of O in GCR increases at late times because their composition results
from a mixture  of ISM and stellar winds; the latter is quasi-constant in time (hence the flat early Be/Fe ratio), while
the former increases steadily and dominates Be production at late times. This enhanced late Be production rate 
 compensates for the late increase
in Fe from SNIa and makes Be/Fe roughly constant also at high metallicities. For that reason [Be/O] (bottom right panel in Fig. \ref{Fig:Be-evol})
displays a behaviour that mirrors the one of [O/Fe]   (top right panel in Fig. \ref{Fig:Be-evol}).
 Because of the large scatter in the O data, it is unclear at present
 whether Be behaves as primary or secondary with respect to oxygen. However, in the stars with the lowest O metallicities
detected so far, the behaviour of Be appears to be much closer to that of a primary element.

%%%%%   HERE: NEW PARAGRAPH
 Despite its success, the simple picture presented here for the evolution of Be cannot be the whole story.
It has been known for some time (Nissen and Schuster 1997) that  [$\alpha$/Fe] displays a bi-modal
distribution in halo stars, and this was recently confirmed by Nisssen and Schuster (2010) with a precise abundance
analysis of 94  stars in the solar neighborhood.  Building on that work, Primas (2010) 
 found systematic differences in the Be abundances among the high  [$\alpha$/Fe] and the low
[$\alpha$/Fe] populations, and her work was substantiated by
Tan and Zhao (2011) with high-resolution and high
signal/noise ratio VLT spectra:  Be abundances in stars with low [$\alpha$/Fe]  ratios
are systematically lower (by 0.3 dex) than in stars with higher [$\alpha$/Fe] ratios, for stars of similar metallicities
(in the range [Fe/H]$\sim$ -1.2 to -0.8). This variation cannot be reproduced in the framework of the simple
model presented in Fig. \ref{Fig:Be-evol}, which is only meant to reproduce average trends of abundances and abundance
ratios. 

At this point it should be recalled that according to the current paradigm of galaxy formation, the Milky Way as a whole and its halo
in particular were formed not through a monolithic collapse, but from the merging of smaller units with different evolutionary
histories. Prantzos (2008) has shown semi-analytically that one key property of the halo, namely,
its metallicity distribution, can be satisfactorily reproduced in that framework, assuming that the smaller - and more abundant -
units had a lower effective yield (attributed to an easier escape of the supernova ejecta from lower potential wells).  
It is expected therefore that the different evolutionary histories of the merging units will affect the chemical evolution of
the halo, producing e.g. some scatter in abundance ratios (Prantzos 2006c).

\begin{figure}
\begin{center}
\includegraphics[width=0.49\textwidth]{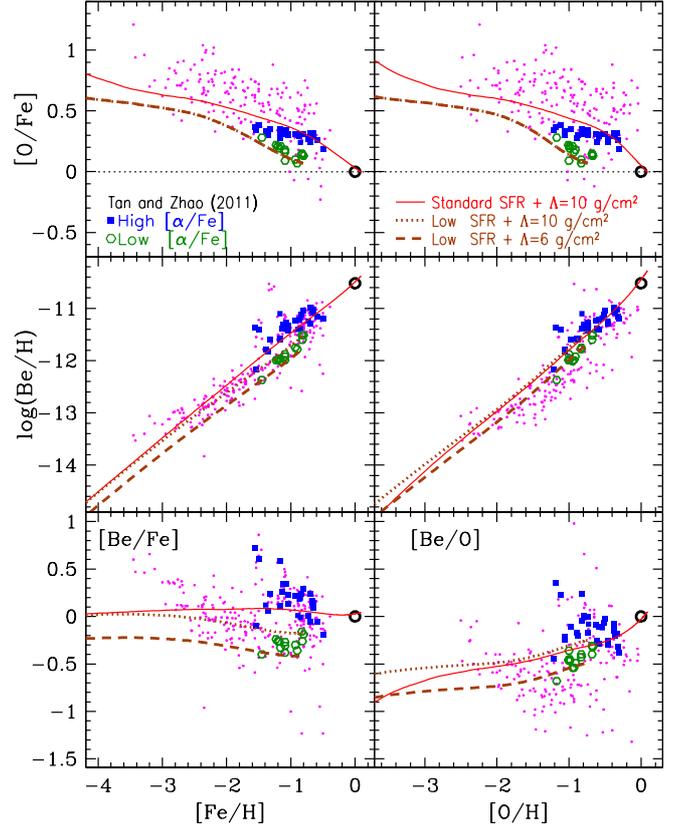}
\caption[]{Same as Fig. \ref{Fig:Be-evol}, with two new data sets ({\it thick symbols})
for stars with high [$\alpha$/Fe] ({\it filled squares}) and low [$\alpha$/Fe] ({\it open squares})
from Tan and Zhao (2011); all other data of Fig. \ref{Fig:Be-evol} are displayed as {\it dots}.
The {\it thin solid} curves correspond to the "standard" model presented in Fig. \ref{Fig:Be-evol},
while the {\it thick dotted} curves to a model with reduced SF efficiency and the {\it thick dashed}
curves to a model with reduced SF efficiency and reduced escape length $\Lambda$ for GCR (see text). 
}
\label{Fig:Be-evol2}
\end{center}
\end{figure}

We perform here a limited investigation of those ideas for the case of Be, aiming to reproduce the findings of Tan and Zhao (2011).
We used oxygen as a proxy for $\alpha$ elements, since they display a similar behaviour for metallicities [Fe/H]$<$-1.5.
The results appear in Fig. \ref{Fig:Be-evol2}. It is seen that the "baseline" model (presented in Fig. \ref{Fig:Be-evol}) reproduces
the data for the high [$\alpha$/Fe]  sample. Adopting a lower SFR efficiency makes it possible to obtain  the observed 
low [$\alpha$/Fe]  values for the same [Fe/H] (upper panels); the reason is that Fe contribution from SNIa comes at lower metallicities
in that case. However, the corresponding low values of Be/O cannot be satisfactorily reproduced, unless it is assumed that the
GCR nucleosynthesis was also less efficient than in the standard case; we chose to simulate this effect by adopting a shorter
escape length ($\Lambda$=6 g cm$^{-2}$ instead of 10 g cm$^{-2}$) and this allows us to obtain  lower Be/O values than in the standard
case.

The expected variety in the physical properties of the merging components of the Galactic halo (size, gas content, potential well,
magnetic field) implies a corresponding variety in the properties of the accelerated particles (different confinement
times and escape lengths). This implies different efficiencies in the spallogenic production of LiBeB, even if the
composition of GCR is assumed to be approximately constant during that period. One should expect
then a rather large - albeit difficult to quantify - scatter in the Be/Fe or Be/O abundance ratios at a given metallicity.  
Incidentally, this argues against  the idea of "Be as a chronometer" (see Smilijanic et al. 2009) and references therein: 
at any given time - or metallicity -
the Be/H value is expected to differ among the evolving components that will later merge to form  the halo. Be would be
a good chronometer only if GCR would have the same properties across the whole galactic system. This excludes the
MW halo (made from components differing in their GCR properties), but could be realised in the local disc, which apparently 
underwent little merging in its late evolution and which is pervaded by a homogeneous GCR "fluid": 
a well-defined relationship between Be/H and stellar age would be the
expected signature of such a process. However, radial migration of stars in the galactic disc is known to
induce scatter in the age-metallicity relation  (Sellwood and Binney 2002), bringing in the
local volume stars born in different galactocentric radii, with different initial abundances.
We calculate the radial profile of Be/H in the MW disc in Sec. 6 and show that a sizeable gradient of Be/H is expected in the ISM,
steeper than for  oxygen. In that case, radial migration would blur any Be-age relation.

%%%%%

\begin{figure}
\begin{center}
\includegraphics[width=0.49\textwidth]{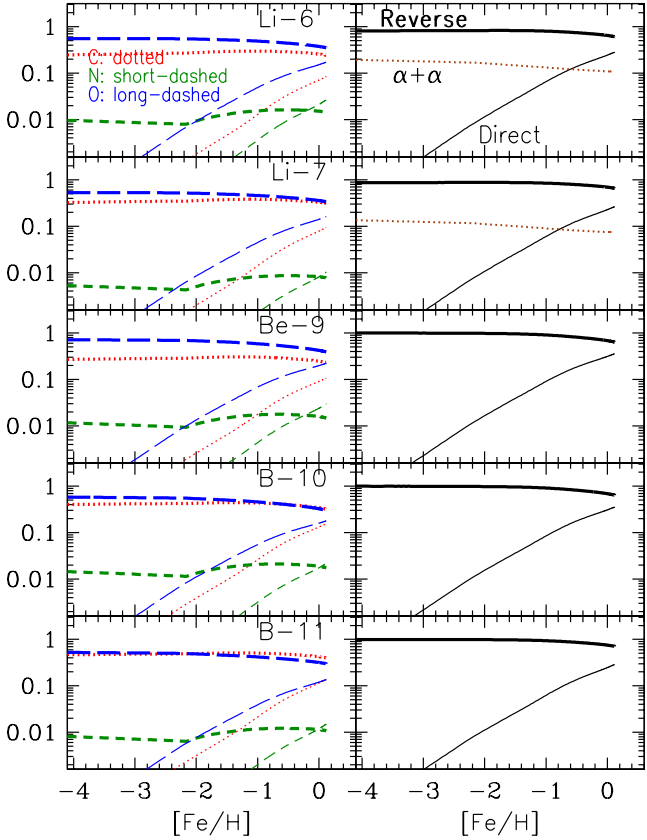}
\caption[]{ {\it Left}: Contributions (percentages)  to the
production rate of Li, Be and B isotopes, with {\it thick curves}
indicating the reverse component and {\it thin} curves the direct one
for C ({\it dotted}), N({\it short-dashed}) and O ({\it long-dashed}).
{\it Right}: Same thing, with the sums of C+N+O contributions for each component
and the contributions of $\alpha$+$\alpha$ ({\it dotted} curves) indicated for the Li isotopes.
}
\label{Fig:LiBeB-components}
\end{center}
\end{figure}

Fig. \ref{Fig:LiBeB-components} displays the detailed contributions of the various components to the
{\it spallogenic} production of Li, Be and B (i.e. percentages of LiBeB production by GCR alone).
In the right panels, it can be seen that the reverse component always dominates LiBeB production, even at
high metallicities; the direct component contributes at most 25-40\% at the highest metallicities. 
This appears counterintuitive, because the reverse component produces fast LiBeB nuclei: some of them are lost
from the Galaxy in the leaky-box model adopted here, while all the slow LiBeB nuclei produced by the direct
component are immediately incorporated in the ISM (the probabilities $P_{ij}^L$ in Eq. 6 are 1 for the direct component terms
but lower than 1 for those of the reverse component). However, this "advantage" of the direct component
is more than compensated for by the higher C and O abundances of the reverse component: 
as discussed in Sec. 3.2, the GCR source abundances of C/H and O/H
adopted here, are considerably higher than those of the ISM. The same is true for the Li isotopes, where
the reverse component dominates even at the lowest metallicities  while the $\alpha+\alpha$ component contributes $\sim$20\%
at most. Because of the enhanced presence of C and O in the reverse component and of its primary nature, 
the situation for Li  is different from that envisioned in Steigman and Walker (1992), who 
suggested that $\alpha+\alpha$ reactions would dominate production of Li isotopes at low metallicities.

In the left panels of Fig. \ref{Fig:LiBeB-components} appear the separate contributions of  C, N, and O to the
spallogenic production of LiBeB.  As expected, C and O dominate, while N has a negligible contribution (at the
1\% level) because of its low abundance. With the adopted GCR composition, O dominates the early production of $^6$Li and $^9$Be, while its contribution matches that of C for the other LiBeB isotopes.

\subsection{Evolution of B isotopes}
\label{Subsec:EvolB}

From the two boron isotopes, $^{10}$B is an almost  100\% product of GCR, like the monoisotopic Be.
Indeed, the meteoritic $^{10}$B/Be ratio is nicely reproduced by our calculations (to better than 10\%) with the adopted
GCR spectra and composition. In fact, the well-known spallation cross-sections (Fig. \ref{Fig:CrossSec}) are the key ingredient here, because  both $^9$Be and $^{10}$B  are produced in about the same amounts by $^{12}$C and $^{16}$O. On the other hand, 
as already mentioned in Sec. 2, a supplementary source of $^{11}$B is required  to obtain the meteoritic
($^{11}$B/$^{10}$B)$_{\odot}$=4 ratio. That source may be the $\nu$-process in CCSN,
 extensively studied in Woosley et al. (1990): a fraction of the most energetic among 
 the $\sim$10$^{59}$ neutrinos  of a SN explosion have energies above a few MeV and are able to 
 spallate $^{12}$C nuclei in the C-shell of the 
 stellar envelope, providing $^{11}$B as well as  some $^7$Li in the He layer (see Sect. 5.1). Soon after the HST observations
 of the primary behaviour of B (Duncan et al. 1992) it was realised that the $\nu$-process can 
 provide such  primary $^{11}$B (Olive et al. 1994). But, if Be is produced as primary by GCR, as suggested by observations, 
 then more than $\sim$50\% of $^{11}$B is also produced as primary by that same process, 
 leaving a rather small role 
 to the $\nu$-process. That role was subsequently investigated in models with parametrised neutrino spectra
 (e.g. Heger et al. 2005; Nakamura et al. 2010). In fact,  the large uncertainties in the  $\nu$ 
 yields of $^{11}$B do not allow one to perform an accurate evaluation of the B evolution: 
 instead the observed B evolution (resulting
 from both GCR and $\nu$-process) has to be used  to constrain the $^{11}$B yields of CCSN. Thus, Yoshida et al. (2008)
 argued that the temperature of $\nu_{\mu,\tau^-}$ and $\overline{\nu}_{\mu,\tau^-}$ 
 neutrinos inferred from the supernova contribution of $^{11}$B in Galactic chemical 
 evolution models is constrained to the 4.3-6.5 MeV range. 
 Notice that the $\nu$-yields of $^{11}$B  depend also on other factors:
 the available amount of
 $^{12}$C in the C-shell, which in turn depends - among other things - on 
 3-$\alpha$ and $^{12}$C($\alpha,\gamma$) reaction rates (see Austin et al. 2011);
 and the compactness of the exploding star, which enhances the neutrino flux 
 (see Nakamura et al. 2010 for progenitor stars of type Ic supernova).

\begin{figure}
\begin{center}
\includegraphics[width=0.49\textwidth]{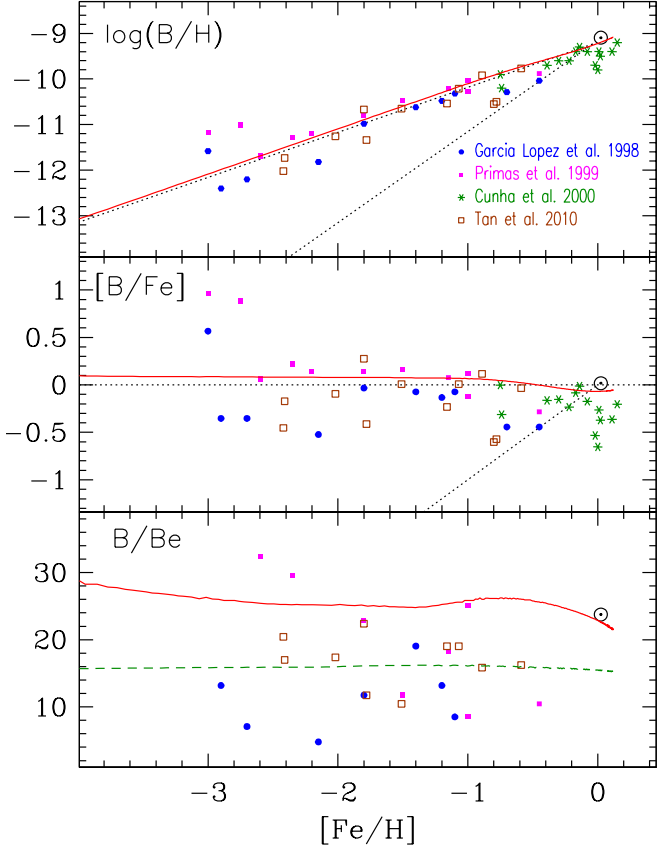}
\caption[]{ From top to bottom: evolution of B/H, B/Fe and B/Be.
In the first two panels {\it dotted lines} indicate primary and secondary evolution. In the bottom panel, 
the {\it solid} curve corresponds to the total $^{11}$B production (GCR + $\nu$-nucleosynthesis) 
and the {\it dotted} curve to $^{11}$B produced by
GCR alone. Data are from: Primas et al. (1999, {\it filled squares}), Garcia-Lopez et al. (1999, {\it filled circles}), 
Cunha et al. (2000,{\it asterisks}), and Tan et al. (2010, {\it open squares}).  
}
\label{Fig:BBe-evol}
\end{center}
\end{figure}

 In Fig. \ref{Fig:BBe-evol} we present the results of our model for the total B (produced by both GCR and $\nu$-process)
 and we compare them to observations. B clearly behaves  as primary with respect to Fe, 
 for the same reasons as Be (see Sec.  4.1). Notice that   to fit the meteoritic B abundance, 
 the $\nu$ yields of $^{11}$B in   WW95 had to be divided by a factor of $\sim$5, otherwise B/H and B/Fe
  would be largely overproduced\footnote{Practioners in the field should be cautious with the use of
  the WW95 yields. It is well known that  to fit the $\alpha$/Fe ratio of halo stars, 
  the WW95 Fe yields should be divided by a factor of two (see e.g. Timmes et al. 1995 or 
  Goswami and Prantzos 2000) and this reduction is adopted here. After considering primary production of Be and B from GCR, the remaining
  $^{11}$B production - to fit the solar $^{11}$B/$^{10}$B - requires  reducing  the
  WW95 $\nu$-yields of $^{11}$B by a factor of 6. If  the
  reduction of Fe yield is not applied, then the reduction of $^{11}$B yields has to be smaller. Furthermore, the WW95 yields at \zs \ are a few times higher than those at  lower metallicities. It is sufficient to 
  apply the reduction factor of 6 only to the solar metallicity yields and not to the others
 to obtain the correct \bb \ ratio at solar system formation; 
 for consistency reasons, we applied that correction to all $^{11}$B yields here.}. 
  Those yields display some metallicity dependence - yields at Z=\zs \ are
  higher than those at Z=0.1 \zs by a factor of a few - and this is visible in the  increase
  of the B/Be ratio after Z=0.1 \zs; however, the late rise of the secondary
  component of Be (exclusively produced by GCR)  makes that ratio decrease again as the
  metallicity approaches Z=\zs. This behaviour is also found in
   the \bb \ ratio (see below).

  Notice that the model B/Be ratio is  $\sim$24 (i.e. approximately solar) during the whole galactic evolution, while the average observed ratio in halo stars is B/Be$\sim$15 and it is compatible with 
  pure GCR production of both elements (dashed curve in the bottom panel of Fig. \ref{Fig:BBe-evol}). 
  However, the large error bars of that ratio prevent  any conclusions and call 
  for future observations  to clarify  this important issue.

\begin{figure}
\begin{center}
\includegraphics[angle=-90,width=0.49\textwidth]{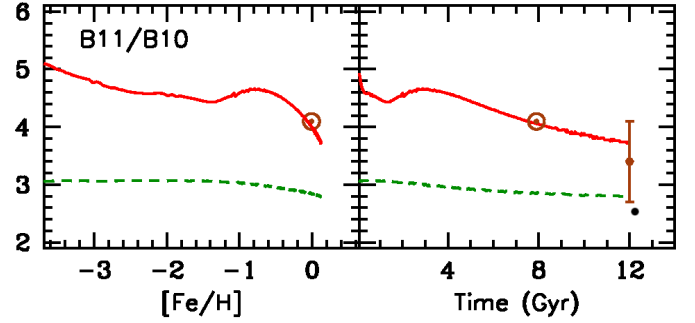}
\caption[]{ Evolution of $^{11}$B/$^{10}$B as a function of metallicity ({\it left}) and time ({\it right}). {\it Solid} curves correspond to the total $^{11}$B production (GCR + $\nu$-nucleosynthesis) 
and {\it dashed} curves to $^{11}$B produced by
GCR alone. The meteoritic (=solar) 
value ($^{11}$B/$^{10}$B)$_{\odot}$=4.02 (Lodders 2003)
is indicated at [Fe/H]=0 and at time $t$=7.5 Gyr, whereas the local
value ($^{11}$B/$^{10}$B)$_0$=3.4$\pm$0.7 (Lambert et al. 1998) is plotted at $t$=12 Gyr. 
}
\label{Fig:Biso-evol}
\end{center}
\end{figure}

Finally, in Fig. \ref{Fig:Biso-evol} we present the results of our calculations for the evolution of
the boron isotopic ratio  $^{11}$B/$^{10}$B. We compare them to the meteoritic ratio 
($^{11}$B/$^{10}$B)$_{\odot}$=4.02
(Lodders 2003) and to the one measured in local diffuse intestellar clouds 
($^{11}$B/$^{10}$B)$_0$=3.4$\pm$0.7 (Lambert et al. 1998). 
If the spallogenic production alone is considered (dashed curves in Fig. \ref{Fig:Biso-evol}), one sees 
that $^{11}$B/$^{10}$B remains
quasi-constant with metallicity (or time) and decreases only slightly at late times. The reason lies in the evolution of the GCR composition
calculated in Fig. \ref{Fig:GCR_CNOcomp} (bottom) and in the weight that C and O have in the production
of the two boron isotopes (Fig. \ref{Fig:LiBeB-components}). 
%At high metallicities (or late times), O increases more rapidly than C in GCR, according to our scheme and to the wind yields adopted here (Fig. \ref{Fig:GCR_CNOcomp}, bottom). 
%Oxygen contributes by a larger amount to  the production of $^{10}$B than to the production
%of  $^{11}$B (Fig. \ref{Fig:LiBeB-components}, bottom left panels). As a result, the production of ${10}$B is favored (with respect to the one of $^{11}$B) at high metallicities and the ($^{11}$B/$^{10}$B)$_{GCR}$ ratio declines. 
At the time of solar system formation ($t$=7.5 Gyr) we find a value of 
($^{11}$B/$^{10}$B)$_{GCR}$=2.8, slightly higher than the usually quoted value of 2.5; this small
difference is attributed to the GCR composition adopted here, particularly enriched in C (which favours
production of $^{11}$B rather than $^{10}$B). 

The evolution of the total ratio ($^{11}$B/$^{10}$B)$_{GCR+\nu}$ (i.e. considering the
production of $^{11}$B by both GCR and $\nu$-nucleosynthesis) is displayed as solid curves in  Fig. 
\ref{Fig:Biso-evol}. The pre-solar (meteoritic) value of (\bb)$_{\odot}$=4 is correctly reproduced by construction,
since the $\nu$ yields of CCSN have been adjusted to that. An interesting feature is the slow decrease of the
predicted \bb \ ratio during the late evolution, for [Fe/H]$>$-0.6, i.e. later than 4 Gyr.
This is a generic feature of the calculation and arises because during this late
evolution, the rising secondary (direct) GCR component  contributes more to $^{10}$B than to $^{11}$B
(which receives a metallicity-independent 40\% contribution from $\nu$-nucleosynthesis);
as a result, the \bb \ ratio declines slowly. Observations (Lambert et al. 1998) find a local
\bb \ ratio lower than, but certainly compatible with, solar, because of large associated uncertainties. 
In any case, according to the  theoretical framework presented in this work, the present-day \bb \ ratio has to be lower than solar.

\section{Evolution of Li isotopes}
\label{Sec:Lievol}

Among the 92 naturally occuring elements, Li is certainly the one with the richest and most complex
history, which is poorly understood at present. The reason is that Li - in particular the isotope
$^7$Li - has three different nucleosynthesis
sites: primordial nucleosynthesis, stars, and GCR. Only the contribution of the latter is relatively well
known at present, because it is tightly connected to the production of Be
 (an exclusive product of GCR) through the corresponding spallation cross-sections (Fig. \ref{Fig:CrossSec}).

The primordial component is uncertain at present, because of the as yet unsettled question of the difference 
between theory and observations: 
the so-called "Spite plateau" of Li/H in low-metallicity halo stars (Spite and Spite 1982)
lies a factor of $\sim$3 below theoretical predictions of standard Big Bang nucleosynthesis (SBBN in the following)
corresponding to the cosmic baryon density provided by WMAP results (see Steigman 2010 and Iocco et al. 2009 for 
recent summaries). To make the situation  worse,
it appears that below [Fe/H]=-2.5 Li/H decreases with decreasing Fe/H and displays some dispersion (Sbordone et al. 2010), two features
that do not characterise halo stars of higher [Fe/H].
 
Mechanisms for Li destruction involving physics beyond the Standard model have been proposed in the
literature (see Jedamzik and Pospelov 2009 and references therein). Astrophysical mechanisms, 
such as astration of high primordial Li  
by a pre-galactic Pop. III population of massive stars (Piau et al. 2006)
face severe problems of metal overproduction (Prantzos 2006c). Alternatively, primordial Li may have been
depleted in the surface layers of halo stars by internal stellar processes (atomic diffusion and mixing
and/or rotation; observational arguments for the latter
alternative have been provided by Korn et al. (2006), on the basis of parametrised models of Richard et al. (2005),
 and their results have been confirmed with a substantially larger sample by Lind et al. (2009). 
In the following we shall adopt the high primordial $^7$Li value, assuming  that its difference with
the Spite plateau is caused by as  yet unspecified  internal stellar processes.

\subsection{Evolution of the stellar component of Li}
\label{Subsec: StellarLi7}

The stellar source of Li is extremely controversial at present\footnote{See contributions to the recent IAU
Symposium 268 "Light elements in
the Universe", Eds. C. Charbonnel et al. (2010).}, but it  involves generical 
production by the $^4$He+$^3$He reaction in various sites:  in AGB stars, 
where the so-called "Cameron-Fowler mechanism" operates in the hot base of their convective envelopes (d'Antona and Ventura 2010 and references therein); in low-mass red giants (RG), through extra deep mixing and the associated "cool-bottom processing" (Sackmann and Boothroyd 1999);
in novae, with explosive nucleosynthesis  in the He-layer accreted onto the white dwarf (Hernanz et al. 1996); and in
core-collapse SN, with the $\mu$ and $\tau$ neutrinos of the explosion producing $^3$He through 
excitation and subsequent de-excitation of the $\alpha$ particles in the He shell (Woosley et al. 1990). 
The various uncertainties still
hampering our understanding of all those sites render the calculated Li yields highly speculative at
present. In the case of RGs and AGBs neither the mass range of Li producers, nor the corresponding Li yields or their possible metallicity dependence
are known.  In the case of CCSN, the temperature of the neutrinosphere - which fixes the energy of spallating neutrinos -  
is poorly known (see Mueller et al. 2012), while the position of the He-layer  - affecting the neutrino flux through the 1/$r^2$ factor - 
depends a lot on poorly constrained stellar physics. Finally, in the case of  novae, neither the Li yields nor the nova rate and its evolution with time are well known.

\begin{table}[!]
\caption{Models of Li production}
\begin {center}
\begin{tabular}{lcc}
\hline \hline 
 \noalign{\smallskip}
 
    & Model A & Model B  \\
    \hline 
 \multicolumn{1}{l}{Primordial} & \multicolumn{2}{c}{SBBN + WMAP}  \\
 
 \multicolumn{1}{l}{GCR} & \multicolumn{2}{c}{Standard GCR (constrained by $^9$Be)}  \\
 
  Massive stars & $\nu$-yieds from WW95  & None \\
  
  Int. mass stars & 3-4 \ms (or novae) & 1-2 \ms  \\
\hline \hline
\end{tabular}
\end{center}

\label{Tab:ModelsLi}
\end{table}

To make things worse, observed Li abundances on the surfaces of low-mass long-lived stars offer no  real constraints on the Li evolution,
contrary to the case of other elements, because Li is depleted in them by poorly known factors: the Spite plateau
most probably does not reflect the true value of Li in the gas from which the stars were formed and the same probably holds  for
disc stars (see Lambert and Berry 2004 for discussion).

\begin{figure}
\begin{center}
\includegraphics[width=0.49\textwidth]{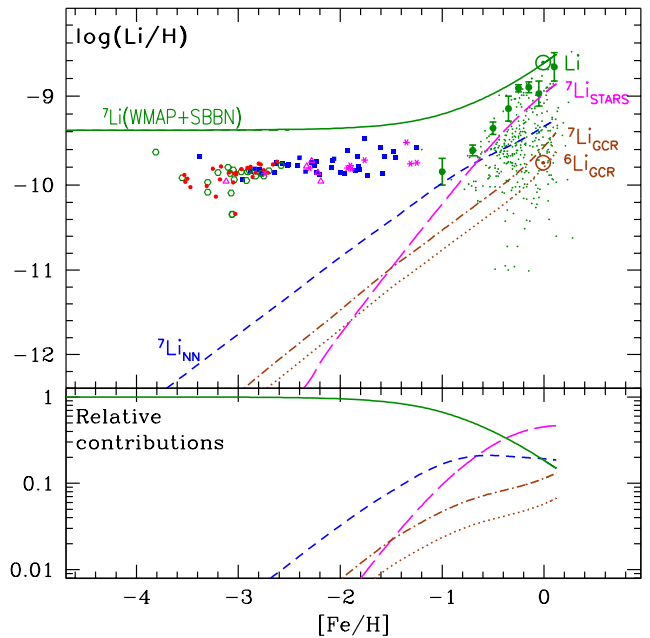}
\caption[]{Evolution of  Li ({\it top}) according to our Model A (see Table 1) and percentages of its various components ({\it bottom}):
$^7$Li  from GCR ({\it dot-dashed}), $^6$Li from GCR ({\it dotted}), $^7$Li  from $\nu$-nucleosynthesis 
(NN, {\it dashed}) and $^7$Li  from a delayed stellar source 
(novae and/or AGB stars, {\it long dashed}). {\it Solid} curves indicate total Li ({\it upper} panel)
and primordial $^7$Li ({\it lower} panel). Abundance data ({\it upper panel} for halo
stars  are taken from Charbonnel and Primas (2005, {\it filled squares}), Sbordone et al. (2010, {\it filled circles}), Bonifacio et al. (2007, {\it open circles}),
Garcia-Perez et al. (2009, {\it open triangles}),  Asplund et al. (2006, {\it asterisks})  
and for disc stars from   Lambert and Reddy (2004, $dots$). In the latter case, points with error bars indicate the
average values of the six most Li-rich stars in the corresponding metallicity bins.
}
\label{Fig:Li-evol1}
\end{center}
\end{figure}

\begin{figure}
\begin{center}
\includegraphics[width=0.49\textwidth]{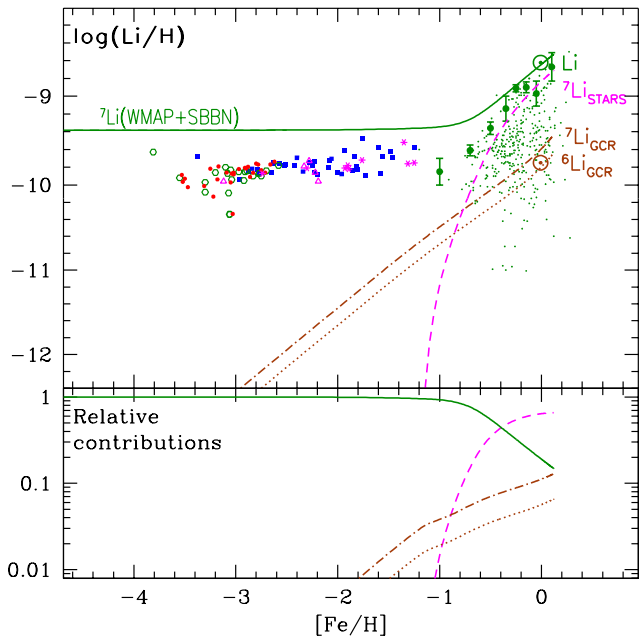}
\caption[]{Same as the previous figure, but for Model B, with no contribution from CCSN and a different assumption for the
stellar Li component  (see Table 1 and text).
}
\label{Fig:Li-evol2}
\end{center}
\end{figure}

In those conditions, we perform calculations for only a few cases - out of the many possible - to illustrate some, 
hopefully realistic,  aspects of the true Li evolution.
For that purpose we adopt Li from four different types of sourcces:

1) SBBN with a high primordial value: log(Li/H)$_P$=-9.4.  

2) GCR, as presented in Sec. 3  (but see next section for the possibility of pre-galactic
  contribution to $^7$Li).

3) $\nu$-nucleosynthesis in CCSN, with yields provided by WW95. 
We keep the nominal values of those yields, without applying a reduction by a factor
of $\sim$6 (contrary to what was done in the case of $^{11}$B yields, 
where boron overproduction had to be avoided).
Although our procedure is not self-consistent, we seek here  the maximum possible contribution 
of the WW95 yields to the Li abundance, to see if they violate any observational
constraint (see also Matteucci et al. 1995); 
results can be easily scaled downwards and even down to zero contribution from CCSN.

4) Stellar sources other than CCSN, that is low-mass RGs, AGB stars, and novae. 
The situation has not evolved much since the extensive discussions 
 in Travaglio et al. (2001) and Romano et al. (2001), which  clearly show that
the uncertainties in all those Li sources  make any  quantitative calculation of Li evolution almost meaningless.

 Our baseline model (Model A) includes all classes of sources (1), (2), (3), and (4). We find that the first three sources
 can produce at most $\sim$45\% of solar (=meteoritic)  Li after 7.5 Gyr of evolution, thus leaving more than half
 of solar Li to be produced by class (4) sources. In Model A we adopt
 2-4 \ms \ AGB stars, their yields 
 (assumed constant with metallicity and stellar mass) being adjusted  to have the solar (=meteoritic)  Li reproduced.
Notice that the death rate of 2-4 \ms \ stars turns out to be approximately 
proportional to that of SNIa, at least according to the formalism of Greggio (2005) for single-degenerate white dwarfs that
is adopted here. In its turn, the evolution of the SNIa rate is as good a guess as can be made at present for the evolution of
the unknown nova rate, because  both phenomena involve accretion onto white dwarfs in binary systems. Then 
the absolute rate of nova is  obtained by    normalising the model value of SNIa rate 
at $t$=12 Gyr by the factor $f=R_{nova}/R_{SNIa}$, where 
$R_{nova}\sim$20-30 yr$^{-1}$ is the   present-day nova rate in the whole Galaxy 
 and $R_{SNIa}\sim$4 10$^{-3}$ yr$^{-1}$
 is the present-day SNIa rate (see e.g. Prantzos et al. 2011 and references therein). 
Thus,
we can use Model A to infer average Li yields of either AGBs of 2-4 \ms \ or of typical novae (assuming that each one of those
sources is, alternatively, the only stellar source of Li of class 4). 
 
 In Model B, we drop source (3), because CCSN can produce at most 20 \% of solar Li and the results of the 
 B evolution suggest a severe downwards revision for the corresponding WW95 yields. We explore then the
 constraints on the Li yields of stars of either 1-2 \ms, 2-4 \ms \  or 2-6 \ms, assuming that each stellar class 
 produces half of  the solar Li. 
 The former class of stars corresponds to the low-mass red giants where Li may be produced by "cool-bottom
processing" according to Sackmann and Boothroyd (1999). The other two  include  most of the AGB stars.

   The results of Model A  appear in Fig. \ref{Fig:Li-evol1}. As already discussed, observations provide no
   constraints on Li/H evolution, because it is assumed that Li in stellar surfaces has been depleted by internal
   processes from its original value.  The first three of the aforementioned
   Li sources can produce a total of $\sim$47\% of the solar 
   Li\footnote{Notice that the model accounts for astration of 
     Li by low-mass stars, but there is also infall of primordial Li during the evolution.}. 
   This leaves about 50\% of the solar Li to    be produced by the low-mass stellar sources. Notice that if the
   CCSN Li yields of WW95 are reduced by e.g. a factor of 6 (as we did for $^{11}$B yields, see Sec. 4.2),
   the CCSN contribution to the solar Li will be reduced by a similar factor (from 18\% to 3\%) and the
   contribution of low-mass sources will grow  by that same amount (15\%), i.e. it will increase from $\sim$50\%
   to $\sim$65\%. 

We find that to produce    $\sim$50\%    to $\sim$65\% of solar Li, 2-4 \ms \ AGBs have to
 eject  a Li yield of $y_{2-4}$(Li)$\sim$ 2 10$^{-7}$ \ms \ on average
(independent of mass or metallicity). These  values are considerably
 higher than most of the Li yields found for  AGB stars by 
 Ventura and d'Antona (as provided in Romano et al. 2001) or by
 Karakas (2010)  for various parametrisations of the AGB envelopes. 
  On the other hand, if novae provide $\sim$50\%    to $\sim$65\% of solar Li, they ought to have a typical
  Li yield of $y_{nova}$(Li)$\sim$ 10$^{-9}$ \ms. This is considerably higher than values found
    with hydrodynamical models by Hernanz et al. (1996) for classical CO novae and, taken at face value,
    it implies that those objects
     are quite insignificant sources of Li in the Galaxy. This conclusion was also reached 
    in Romano et al. (2001)     on the basis of similar arguments.

\begin{table}
\caption{Yields of Li (in \ms) from low-mass sources}
\begin {center}
\begin{tabular}{lcr}
\hline \hline 
 \noalign{\smallskip}
 
    & Inferred  from & Calculated  \\
       &  this work$^a$ &  in literature (Ref.)  \\
\hline 
 Novae  &    10$^{-9}$  & 10$^{-10}$  (1)    \\
 AGB (2-6 \ms)  &    1.5 10$^{-7}$  & $<$ 3. 10$^{-8}$  (2, 3)   \\
 AGB (2-4 \ms)  &    2.  10$^{-7}$  & $<$ 5 10$^{-9}$  (2)   \\
 RG (1-2 \ms)  &    1.  10$^{-7}$  &  $<$ 10$^{-8}$ (3) \\
\hline \hline
\end{tabular}
\end{center}
$a$: Under the assumption that each class of sources produces 50\% of solar Li (the remaining coming from SBBN+GCR+$\nu$-nucleosynthesis in CCSN).
$References$: (1): Hernanz et al. (1996), (2): Ventura and d'Antona, yields presented in Romano et al. (2001) and (3): Travaglio et al. (2001).

\label{Tab:YieldsLi}
\end{table}

 The results of Model B  appear in Fig. \ref{Fig:Li-evol2}. CCSN are dropped as Li sources, and 1-2 \ms \ stars 
 start
 enriching the ISM at [Fe/H]$\sim$-1, i.e. about 1 Gyr after the beginning. To produce the
 required 70\% of solar Li (SBBN and GCR accounting for $\sim$30\%), these low-mass stars must produce
 an average  yield of $y_{1-2}$(Li)$\sim$  10$^{-7}$ \ms. Again, these values are much higher than those evaluated
 for red giants by e.g. Travaglio et al. (2001) or 
 Karakas (2010). Notice that the sharp rise with metallicity of the Li contribution from
 low-mass stars with metallicity-independent Li yields mimics the behaviour of higher mass stars with metallicity-dependent
 Li yields (increasing with increasing metallicity). Even in that case, however, the requirement of Table 2 on the $average $ yields
 remain.

 It seems then that the main Li source remains elusive at present: clearly, the firmly
 established Li sources (SBBN and GCR) can only produce $\sim$30\% of the solar Li and the uncertain contribution
 of $\nu$-nucleosynthesis from CCSN yields at most another $\sim$20\%; but the 
 remaining part - the majority of solar Li - can hardly be explained
 by present-day models of novae, RGs or AGB stars.

 The situation may appear more optimistic in the case of novae: 1D hydrodynamical
  calculations of nucleosynthesis in CO novae
 find an overproduction factor of $\sim$10$^3$ for \lia \ (Hernanz et al. 1996) in the ejecta, i.e. a mass fraction
 of $X_7\sim$10$^{-5}$, and a typical  ejecta mass of  $\sim$10$^{-5}$ \ms; the latter, however, is $\sim$5-10
 times lower than {\it  observationally inferred} ejecta masses, suggesting a problem of 1D nova models at this level.  
 If this "low ejecta mass" problem is fixed, then for the same overproduction factor the \lia \ yield
 will increase to 10$^{-9}$ \ms, exactly to the level required by chemical evolution arguments. In that case, 
 our baseline Model A would be, perhaps, not far from reality - provided  the nova rate is approximately
 proportional to the SNIa rate and the SNIa rate follows the Greggio (2005) formalism, as assumed here.
 However, as Travaglio et al. (2001) correctly point out, in that case novae would overproduce by large factors
 the minor CNO isotopes ($^{13}$C, $^{15}$N and $^{17}$O, see their Table 2); 
 this constitutes a powerful argument
 against novae as significant \lia \ producers and leaves RGs and AGBs as the only viable source. 
 It should be stressed, at this point, that nucleosynthesis in those sites
 has been  calculated with 1D models up to now and that mass loss plays a critical role
 in the overall evolution and the \lia \ yields of those sources.  
 The introduction of 2D or 3D models  and a better treatment of mass loss may change 
  our understanding
 of those complex astrophysical sites considerably and, perhaps, increase  their Li yields drastically.

\begin{figure}
\begin{center}
\includegraphics[width=0.49\textwidth]{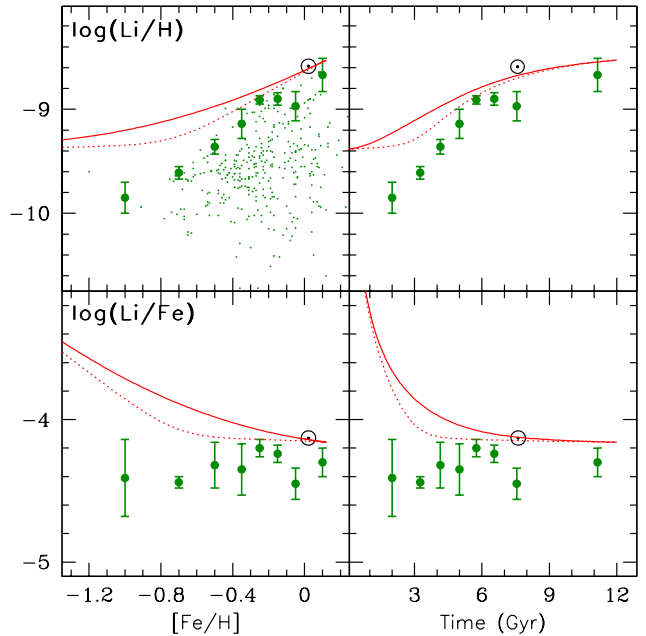}
\caption[]{Evolution of  Li/H ({\it top}) and of Li/Fe ({\it bottom}) in the local disc, as a function of [Fe/H] ({\it left}) and
of time ({\it right}). In all panels,   {\it solid} curves correspond to Model A and   {\it dotted}  curves to Model B.   Data 
 are from Lambert and Reddy (2004) and are provided as a function of metallicity. They  
are plotted as a function of time ({\it right} panels) by applying  the model age-metallicity relation.
}
\label{Fig:Li-disc}
\end{center}
\end{figure}

 In Fig. \ref{Fig:Li-disc} we plot the late evolution of Li for metallicities higher than [Fe/H]=-1.3
 and we compare it with data for the local disc from Lambert and Reddy (2004). In their discussion, Lambert and Reddy (2004) acknowledged the possibility that the upper envelope of their data - represented
 by the averages of the six most Li-rich stars in each of their [Fe/H] bins -  does not reflect the true
 Li evolution, contrary to views held previously (before the release of WMAP results). After a detailed analysis
 of the Li abundances as a function of the metallicities and masses of the stars of their sample,  Lambert and Reddy (2004)
 suggest that, at least for [Fe/H]$<$-0.4, the true Li/H value was probably higher by $\sim$0.5 dex and, therefore,
 Li/H may have evolved little through the lifetime of the thin disc.  Our results support this view, although they are clearly model-dependent: different choices of the rate of the main Li source, or even simply the adoption
 of metallicity-dependent Li yields (increasing with metallicity) would lead to a  model curve closer to  the data. 
 Taken at face value, the difference between our model curve and the data may imply a metallicity- (or age-)
 dependent depletion of Li in stars. However, as Lambert and Reddy (2004) pointed out, other factors, such as
 mass and rotation may also play a role.  Clearly, a lot of work is still required to separate 
 the various factors affecting the evolution of Li inside stars and in the ISM.

\begin{table}
\caption{Contributions (\%) of various sources to solar LiBeB production}
\begin {center}
\begin{tabular}{lcccc}
\hline \hline 
 \noalign{\smallskip}
  &  SBBN    &   GCR   &  $\nu$ in CCSN  & Low-mass stars$^a$ \\
\hline 
$^6$Li       &      &  100$^b$   &     &     \\
$^7$Li       &   12   &    18 &  $<$20      &  50-70     \\
$^9$Be       &      & 100    &     &     \\
$^{10}$B    &      &  100   &     &     \\
$^{11}$B    &      &    70  &  30    &     \\
  
\hline \hline
\end{tabular}
\end{center}
$a$: Red giants, AGBs, novae ; $b$: Assuming no pregalactic $^6$Li.

\label{Tab:LiBeBcontr}
\end{table}

Finally, Table \ref{Tab:LiBeBcontr} summarises the contributions of the various sources to the solar
abundance of each one of the LiBeB isotopes.  As already stated, three of those isotopes, namely 
$^6$Li, $^9$Be and  $^{10}$B, are exclusively produced by GCR (with the possible exception of 
a pre-galactic production for  $^6$Li, see next section). $^{11}$B requires the contribution
of $\sim$30\% of a non-GCR source, presumably $\nu$-nucleosynthesis. Finally, $\sim$12\% of the
solar   $^7$Li  is due to SBBN (after accounting for all factors affecting its evolution, namely
astration and infall) and $\sim$20\% to GCR. This leaves $\sim$70\% to the stellar source(s);
$\nu$-nucleosynthesis can produce {\it at most} 20\%, but certainly much less, leaving more than
half of the solar   $^7$Li to be produced by low mass stars.

\subsection{Early  $^6$Li: ``high'' or ``low'' ?} 
\label{Subsec:EarlyLi}

\begin{figure}
\begin{center}
\includegraphics[width=0.49\textwidth]{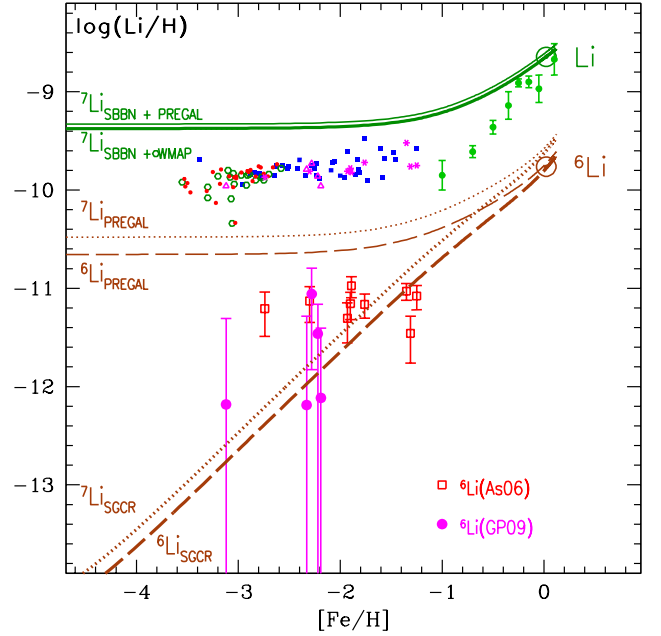}
\caption[]{Evolution of total Li  ({\it solid curves},  $^6$Li ({\it dashed curves}) and   $^7$Li from GCR ({\it dotted curves})   
in the standard case (low \lib \ from SBBN, {\it thick curves})  and in the case of
high pre-galactic \lib \ and \lia \ ({\it thin curves}). 
In the latter case, a minimum amount of depletion of \lib \ within stars (equal to that of \lia) has been conservatively assumed. 
\lia \  data for halo stars  are as in Fig.
\ref{Fig:Li-evol1},  while   \lib \ 
data are from Asplund et al. (2006, As09, {\it open squares}) and Garcia-Perez et al. (2009, GP09, {\it filled circles}).
 
}
\label{Fig:Li6evol}
\end{center}
\end{figure}

The report of an ``upper envelope"  for $^6$Li/H in low-metallicity halo stars by
Asplund et al. (2006) gave a new twist to the 
LiBeB saga. The reported $^6$Li/H value
at   [Fe/H]=-2.7 is much higher - by a factor of fifteen -  than expected if GCR are the only source
of the observed $^6$Li/H in that star, assuming that GCR  account for the observed evolution of Be  
(see Fig. \ref{Fig:Li6evol}). Things are even worse if the true primordial Li is the one corresponding to the WMAP+SBBN value (as assumed here), because $^6$Li is more fragile than $^7$Li: in that case,  the
initial $^6$Li  values in halo stars should be at least a factor of 3 higher than evaluated 
by Asplund et al. (2006), bringing the dicrepancy with the theory  to a factor of $>$40.

Cayrel et al. (2007) argued that asymmetric convective motions in stellar atmospheres could alter the
profile of the Li line, mimicking the presence of \lib. Because of those uncertainties,
the reality of  high $^6$Li values in halo stars has not been definitely established yet; the answer will
require 3D model atmospheres in the non-LTE regime (see Asplund and Lind 201;, Steffen et al. 2010
and Spite and Spite 2010 and references therein). As clearly stated in Asplund and Lind (2010): "it is not yet
possible to say that \lib \ has definitely been detected, but it is definitely too early to say that \lib \ has
not been detected". 
 
The possibility of important pre-galactic production of $^6$Li by non-standard GCR has drawn 
considerable attention from theoreticians, who proposed several scenarios:

1) Primordial, non-standard, production during Big Bang nucleosynthesis:
 the decay/annihilation of some massive particle (e.g.
neutralino) releases energetic nucleons/photons that produce $^3$He or  $^3$H
by spallation/photodisintegration of  $^4$He, while
subsequent fusion reactions between $^4$He and $^3$He or  $^3$H 
create  $^6$Li (e.g.  Kusakabe et al. 2008; Jedamzik and Pospelov 2009 and references therein). 
Observations of  $^6$Li/H constrain then the masses/cross-sections/densities of the massive particle. 

2) Pre-galactic, by fusion reactions of  $^4$He nuclei, accelerated
by  the energy released by massive stars (Reeves 2005) or by shocks induced 
during structure formation (Suzuki and Inoue 2002; Rollinde et al. 2005, 2006; Evoli et al. 2008).

3) In situ production by stellar flares, through $^3$He+$^4$He reactions  (Deliyannis
and Malaney 1995)  involving large amounts of accelerated $^3$He (Tatischeff and Thibaud 2007).

Prantzos (2006b) showed that the energetics of $^6$Li production by accelerated particles severely constrain
any scenario proposed in category (2) above, including jets accelerated by massive black holes; this
holds  also for the ``stellar flare" scenario (3), 
the parameters of which have to be pushed to their extreme values
 to obtain the ``upper envelope" of the Asplund et al. (2006) observations. This difficulty is confirmed
by Evoli et al. (2008), who calculated pre-galactic $^6$Li production by $\alpha+\alpha$ reactions with a semi-analytical 
model for the evolution of the early Milky Way; they found maximum values lower by factors $>$10 (and 
plausible values lower by 3 orders of magnitude) than the values reported by Asplund et al. (2006). 
Acording to the authors "neither the level nor the flatness of the \lib \
 distribution with [Fe/H] can be reproduced under the
most favourable conditions by any model in which \lib \ production is tied to a (data constrained) 
Galactic star formation rate via cosmic ray spallation."
Those results confirm that the putative pre-galactic cosmic rays are unable to produce \lib \ at the level 
suggested by Asplund et al. (2006), unless  they are powered by some new  energy source, unrelated to star formation and Fe production.
Notice also that the problem becomes {\it much worse}  if it is assumed that the enhanced \lib \ abundance is {\it cosmic}
and not {\it local} (see discussion in Prantzos 2006b): in the former case, the total baryonic content of the Universe is polluted
to that high level, while in the latter only the material involved in the formation of halo stars is concerned. Taking into account
that the mass of the halo is $\sim$1\% of the Milky Way and that the baryons in stars or intracluster plasma today constitute
only $\sim$10\% of the total baryonic content of the Universe (90\% being in intergalactic plasma, Fukugita and Peebles 2004),
ones sees that the energetic requirements are 1000 times more severe in the former case than in the latter.

In Fig. \ref{Fig:Li6evol} we plot the results of our model for \lib, assuming it is exclusively produced
by standard GCR (thick  curve) and a pre-galactic production from some unspecified mechanism (thin curve). In
the latter case, it is assumed that the pre-galactic \lib \ value exceeds the "Asplund upper envelope" by
a factor of three, i.e. by the same factor separating the SBBN+WMAP value from the Spite plateau. 
This a conservative estimate of \lib \ destruction within stellar envelopes, because \lib \ is more fragile
than \lia \ and it should be more depleted than the latter. The total pre-galactic Li (\lia+\lib) produced
in that case by cosmic rays  is about 17\% of the Li produced in SBBN(+WMAP). It seems implausible that such an energetically inefficient process  as spallation-fusion reactions can produce almost as much Li as the Big Bang itself. If it turns out that
$\sim$5\% of Li in halo stars is indeed in the form of \lib \ as claimed by Asplund et al. (2006), a solution involving localised
processes (rather than affecting the total baryonic content of the Universe), 
should be favoured,  because it is less problematic energetically.

\subsection{Evolution of the \lis \ ratio}
\label{Subsec:Li6/Li7}

After the determinations of the \lis \ isotopic ratio in the local ISM in the 1990ies (Lemoine et al. 1993; Meyer et al.
1993),  Reeves (1993) and Steigman (1993) assessed the importance of that ratio   for our understanding of Li sources and chemical evolution since the formation of the Sun. However, the complexity of the topic (related
to the large number of unknowns) made a quantitative - and even a qualitative - assessment impossible\footnote{
For instance, at that time the dominant role of the reverse (primary) GCR component to the production of \lib \ 
at all metallicities was not realised.}.

In this section we reassess the late evolution of the \lis \ ratio on the basis of our models and motivated
 by the recent measurements of that ratio in the local ISM by Kawanomoto et al. (2009).
While Lemoine et al. (1993) found a \lis \ value equal to the solar one along one line of sight, Meyer
et al. (1993) found more than twice that value along two lines of sight. Kawanomoto et al. (2009) found
intermediate values along three lines of sight and their 1$\sigma$ error bars extend between the values
found by Lemoine et al. (1993, 1995) and Meyer et al. (1993). In view of those results, it appears that the
\lis \ ratio in the local ISM has either remained almost constant or increased in the past 4.5 Gyr.

\begin{figure}
\begin{center}
\includegraphics[angle=-90,width=0.49\textwidth]{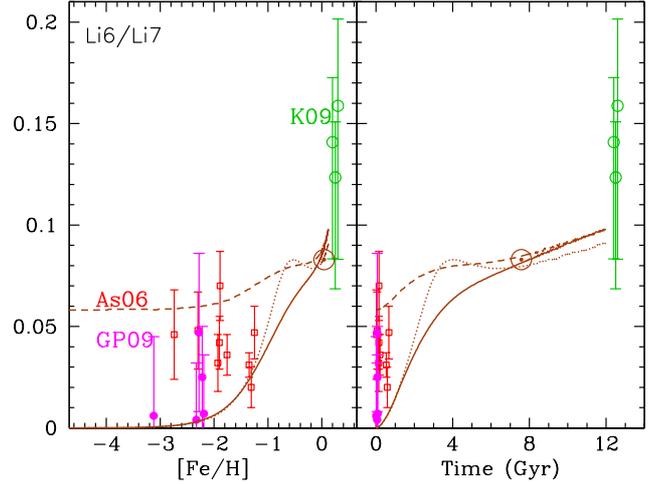}
\caption[]{{\it Left:} Evolution of \lis \ ratio as a function of [Fe/H] ({\it left}) 
and of time ({\it right}). Data are taken from Asplund et al. (2006, As06), Garcia-Perez et al. (2009, GP09)
and Kawanomoto et al. (2009, K09). {\it Solid} curves correspond to Model A and {\it dotted} curves to Model B, respectively,
both starting with  standard (low) pre-galactic $^6$Li; {\it dashed} curves correspond to Model A with high pre-galactic $^6$Li. 
}
\label{Fig:Li6ratio-halo}
\end{center}
\end{figure}

Fig. \ref{Fig:Li6ratio-halo} displays the evolution of the \lis \ ratio according to our models 
compared to data for the early halo
(highly uncertain, see discussion in previous section) and in the local ISM. 
Theoretical predictions for the early evolution obviously depend on the adopted pre-galactic  \lis \ ratio.
Late evolution depends on the adopted yields of \lia, since the evolution of \lib \ is well determined.
A generic feature of our models (i.e. independent of whether the sources are low mass RGs, AGBs or novae)
is the slow late rise of \lis, which brings the model results up to  \lis$\sim$0.1, i.e.
within the 1$\sigma$ error bars
of the measurements. This slow increase is due to the fact that \lia \ behaves essentially as a a primary
(the dominant stellar component has metallicity-independent yields) while \lib \ has a substantial secondary
contribution ($\sim$30\%) from its direct component (thin curve in the right panel of Fig. \ref{Fig:GCR_CNOcomp}).
 The increase of \lib \ is thus slightly faster than the one of \lia \ and this 
 is reflected in the rise of the \lis \ ratio.

Notice, however, that if the \lia \ yields of low-mass sources turn out to be increasing with metallicity,
the tendency would be inversed and the late \lis \ ratio  would be found to decrease (before or
after solar system formation). On the other hand, if the current 1$\sigma$ error bars of the measurements 
shrink to, say, 0.01 (and the average local \lis \ value is still twice as high as the meteoritic one),
then possible alternatives might involve \lia \ yields decreasing with metallicity;
or localised irradiation of the ISM  by GCR, increasing the \lis \ ratio. 
Much more precise measurements of the local \lis \ ratio are required  to definitely conclude
on this important question.

\section{Radial profiles of LiBeB in the MW disc}
\label{Subsec:Radial}

Profiles of chemical abundances in galactic discs provide key diagnostics of the evolutionary processes
that shaped them because they depend on i) stellar nucleosynthesis, i.e. whether they were made as primaries
or secondaries and in short- or long-lived sources,  and ii) galactic processes, in particular the ratio
between the rates of star formation and infall. For the Milky Way disc, much
observational and theoretical work has been made
on the profiles of several key elements, such as oxygen, and their evolution (see e.g. the discussion
in Chapt. 4 of Stasinska et al. 2012 for  oxygen profiles).

The abundances of the light elements Li, Be and B have not been observed in other parts of the
Milky Way disc except in the solar neighbourhood, up to now. Still, it is interesting to see what current
models of the MW disc evolution predict for the radial profiles of those elements and their isotopes,
because future observations may provide additional  constraints on models
of LiBeB evolution, supplementing those already obtained in the local halo and disc.

We adopt here a considerably updated version of the evolutionary model for the Milky Way disc presented
in Boissier and Prantzos (1999), which satisfies all the major observational constraints (radial
profiles of gas, stars and SFR, total rates of CCSN and SNIa, luminosities in various wavelength bands, etc.).
That model was also used to study in considerable detail the radial profiles of several elements,
confronting successfully model predictions to observations (Hou et al. 2000). Among the various improvements 
brought to the
model, which are relevant for this work,  are the introduction of the Greggio (2005) SNIa rate; and, in particular, the implementation of the
full formalism of LiBeB production (Sec. 3.3) taking into account the evolving composition of GCR (according
to Sec. 3.2). The results presented here have been obtained with Model A for Li production (see Sec. 5.1).

\begin{figure}
\begin{center}
\includegraphics[width=0.49\textwidth]{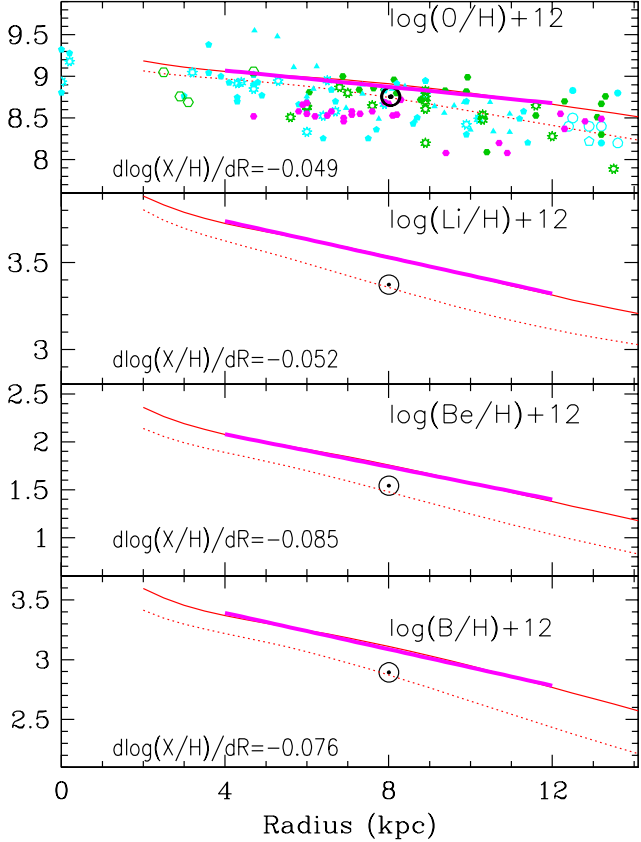}
\caption[]{ Radial profiles of O/H,  Li/H,    Be/H and B/H   (from top to  bottom)
at time $t$=7.5 Gyr (solar system formation,  {\it dotted} curves) and at  $t$=12.Gyr (now,  {\it solid} curves). 
Present-day profiles between 4 and 12 kpc are fitted with exponentials 
({\it thick segments}) and the corresponding slopes appear in the lower left corner
of each panel.
}
\label{Fig:OLiBe_profiles}
\end{center}
\end{figure}

The results of the model for O, Li, Be and B are presented in Fig. \ref{Fig:OLiBe_profiles}. Oxygen presents a 
present-day gradient of dlog(O/H)/dR=-0.049 dex/kpc in the range 4-12 kpc,
in broad agreement with observations (although there
is no general agreement as to the precise value of that gradient and even on the exact shape of the
oxygen profile at present). Li displays a similar profile, clearly suggesting its primary nature,
since it behaves like oxygen. The reason is, obviously, the fact that in Model A the
major Li source is low-mass stars with metallicity-independent yields. Be, on the other hand,  displays an
interestingly different behaviour, since its final abundance profile is considerably steeper
than those of O and Li. The reason is that in the late evolution of Be, the secondary component plays a
substantial role (Fig. \ref{Fig:LiBeB-components}), which becomes even more important in the inner, 
metal-rich regions of the disc. In the case of B, late evolution involves both the secondary component and
the primary one (from $\nu$-nucleosynthesis), making its profile less steep than the one of Be.
Thus, detection of the Li, Be and B profiles in the MW disc could provide crucial
information on the importance of the secondary component on the production of those elements: for instance,
if the Li profile turns out to be much steeper than predicted here, it would imply that the stellar
Li source (be it CCSN, RGs or AGbs) has metallicity-dependent yields. Similarly, a Be profile flatter than
predicted here would imply that the secondary component plays a much less important role than found here,
and this would in turn impact on our assumptions on the GCR composition in the inner disc. Obviously, 
such conclusions would be drawn only after  a proper treatment of various biases, 
such as depletion of Li in stars
by internal processes and in the ISM by fractionation.

\begin{figure}
\begin{center}
\includegraphics[width=0.49\textwidth]{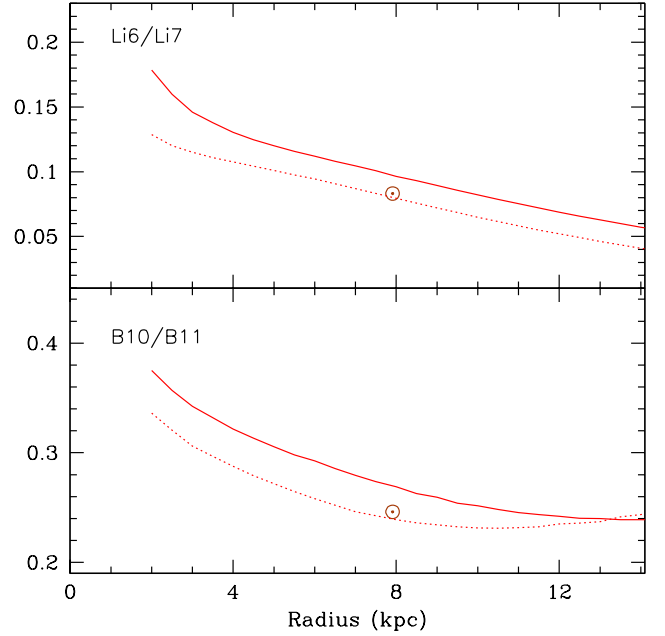}
\caption[]{Radial profiles of \lis \ ({\it top}) and \bb \  ({\it bottom})
at time $t$=7.5 Gyr (solar system formation,  {\it dotted} curves) and at  $t$=12.Gyr (now,  {\it solid} curves).
}
\label{Fig:LiBiso_profiles}
\end{center}
\end{figure}

Finally, in Fig. \ref{Fig:LiBiso_profiles} we present the results for the abundance ratios of the Li and B
isotopic ratios. Both \lis \ and $^{10}$B/$^{11}$B 
increase towards the inner disc, for reasons similar to those discussed
in the previous section for Be: in each one of the two isotopic ratios, the isotopes in the nominator (\lib \ 
and  $^{10}$B, respectively) have a stronger secondary component than those in the denominator (because the
latter also have  substantial primary components). It is precisely for this reason that
\lis \ and  $^{10}$B/$^{11}$B increase slowly in the last few Gyr in the solar neighbourhood 
(see Figs. \ref{Fig:Li6ratio-halo} and \ref{Fig:Biso-evol}, respectively).
This secondary component is more pronounced in the higher metallicities of the inner galactic regions, 
making the corresponding isotopic ratios reach higher 
values there. This is a robust prediction of the model and any qualitative deviations from it - for example,
a flat \bb \ profile - would have profound implications for our understanding of the synthesis of the
corresponding isotopes.

\section{Summary}

In this work we reassessed the problem of the production and evolution of the light elements
Li, Be and B and of their isotopes in the Milky Way in the light of new observational and theoretical
developments.
 The main novelty with respect to the large
body of previous theoretical work in the field is the introduction of a new scheme for the origin
of Galactic cosmic rays (sec. 2.3), which  for the first time enables a self-consistent calculation of their 
evolving composition (Sec. 3.2). The scheme, proposed and quantitatively elaborated in Prantzos (2012), 
accounts for the key feature of the present-day GCR source  composition, namely the high \neo \ ratio, and is based
on the wind yields of the Geneva models of rotating, mass-losing stars (Sec. 2.4). 
The adopted model of Galactic chemical
evolution (Sec. 3.1) satisfies all the major observational constraints in the solar neighbourhood and it is properly
coupled to the detailed formalism of LiBeB  
production (Sec. 3.3) through the energetics of GCRs and of their power sources (CSSN and SNIa).

The GCR source composition is strongly enhanced in CO nuclei,  rendering the reverse LiBeB component
(fast CNO nuclei hitting protons and alphas of the ISM)  dominant at all metallicities.
We  showed that the GCR composition calculated within the new scheme leads naturally to a primary
production of Be, as observed (Sec. 4.1).
Although the result is not new, we think that it is now established - both
qualitatively and quantitatively -
on  more solid theoretical foundations than before (i.e. when incorrect arguments about GCR composition
originating in superbubbles were used). Moreover, the adopted GCR composition helps understanding why
Be follows Fe better  than O: the late increase of C and O in GCR (because of the contribution of the
accelerated ISM) compensates for the late Fe contribution of  SNIa and makes the Be/Fe ratio nearly constant.
We interpreted the recent finding  of lower Be abundances in stars with lower [$\alpha$/Fe]
ratios (Tan and Zhao 2011) in terms of the inhomogeneous chemical evolution of the halo in the framework of hierarchical
galaxy formation. We argued that this implies that Be (or any other element) cannot be used as a "chronometer", at least
for the period of the halo evolution.
 
We found (Sec. 4.2) that GCR alone can produce a boron isotopic ratio of \bb=2.8
at solar system formation, i.e. about 70\% of the meteoritic one; the remaining 30\% of $^{11}$B can
be provided by $\nu$-nucleosynthesis in CCSN, but the WW95 yields of $^{11}$B have to be reduced by a factor
of $\sim$6 to obtain that result. We showed that the \bb \ ratio has to decline slowly in late times, because
of the rising importance of the direct component - producing secondary  LiBeB -, which contributes more to 
$^{10}$B than to $^{11}$B. Current observations are compatible with that trend, but their 
error bars are too large to allow conclusions.

We showed that the two well-known sources of Li, namely primordial nucleosynthesis and GCR, 
can provide $\sim$12\% and $\sim$18\% of solar Li, respectively
(assuming a primordial \lia \ value obtained through SBBN+WMAP).
$\nu$-nucleosynthesis in CCSN can provide another 20\% at most, assuming nominal values for the WW95 yields
of \lia; if those yields are reduced by the same factors as those of $^{11}$B, then CCSN contribute only
$\sim$3\% of solar Li (Sec. 5.1). In any case, more than 50\% of solar Li require another source, 
which must involve low stellar mass objects: RGs, AGBs, and/or novae. The \lia \ yields of all those sources suffer from large
uncertainties, making any quantitative evaluation of their role infeasible at present. Here we inverted the question
and estimated the average \lia \ yield of each one of those candidate sources, 
{\it assuming that it is the sole source of 50\% of solar Li}. It turned out that those yields are higher by
substantial factors ($\sim$10) than yields proposed in the literature (Table 2): the low-mass source of
Li is unknown at present. New models, accounting properly for  mass loss and mixing effects in 2D or 3D, may
help to improve our understanding of Li production in those sites. 

Claims for a high abundance of  \lib \ in low-metallicity halo stars 
remain controversial at present. We confirm that standard GCR can produce only 1/15th 
of the claimed value (Sec. 5.2); if depletion of \lib \ by a factor of $\sim$3 is allowed 
(as is done for Li in the "Spite plateau"), then the discrepancy between theory and observations
rises to a factor of $>$45. Energy requirements put stringent constraints on any sources of pre-galactic
\lib \ production through spallation-fusion reactions, in particular on astrophysical
sources assumed to enrich the total baryonic content of the Universe. 
If the observations of high early \lib \ are confirmed, then only localised astrophysical sources (i.e. within
the galactic structures enriched in \lib) should be considered as viable. Similarly to the case of \bb \ isotopic
ratio, we showed that the \lis \ ratio should slowly increase at late times. 
Again, observations are compatible with that prediction, but large error bars 
make it impossible to draw any quantitative conclusion (Sec. 5.3).

Finally, the radial abundance profiles of LiBeB in the MW disc were calculated, through
 a detailed model that reproduces all the relevant observational constraints (Sec. 6). 
We found that Li has a present-day profile similar to that of O, i.e. it displays a purely
primary behaviour (because low-mass stars, which constitute its major production  component, are assumed here
to have constant \lia \ yields). In contrast, Be is found to display a much steeper profile, because
the direct component of its production by GCR is secondary in nature and becomes important in the inner disc.
For the same reason, the \lis \ and $^{10}$B/$^{11}$B profiles are found to rise in the inner Galaxy. 
We argued that future and accurate observations of those isotopic ratios  are required to understand the
respective roles of the reverse (primary) and direct (secondary) components of LiBeB production, since their
importance becomes comparable only at late times.

In summary, this work provides a coherent theoretical framework allowing one to put in perspective
the vast body of relevant observational data and to study all aspects  of the 
LiBeB production and evolution in the Galaxy. 
The identification of the main Li source and the extent of Li depletion in the envelopes
of low-mass stars remain, in our opinion, the major open questions
in the field, as far as theory is concerned; other important problems are related to the possibility
of pre-galactic \lib \ production,  the extent of $\nu$-nucleosynthesis in CCSN and the GCR properties in
the merging units that formed the halo. 
From the observations point of view, key questions concern: the upper envelope of Li/H across
the full metallicity range, from the earliest halo stars to the youngest stars in the solar neighbourhood;
the amount of early \lib \ and the B/Be ratio in low $Z$ halo stars; the accurate determination of 
\lis \ and \bb \ isotopic ratios in the local ISM; and the determination of radial LiBeB abundance
profiles across the MW disc. 

\medskip
\noindent
{\it Acknoweledgments}: I am grateful to F. Primas and to G. Meynet, R. Hirschi and T. Decressin for
providing data and for enightening discussions; also to Poul Nissen for drawing my attention to the
problem of Be abundances in stars with low and high $\alpha$/Fe ratios and to an anonymous referee for a
constructive report.

\end{document}